\documentclass[manuscript, screen]{acmart}
\usepackage{booktabs} 
\usepackage{lineno,hyperref}
\usepackage[ruled]{algorithm2e}
\usepackage{array}
\usepackage{lscape}
\usepackage{amsmath}
\usepackage{listings}
\usepackage{hhline}
\usepackage{color}
\usepackage{calc}
\usepackage[normalem]{ulem}

\usepackage{amssymb}
\usepackage{graphicx}
\usepackage{tikz}
\usepackage{rotating}

\def\checkmark{\tikz\fill[scale=0.4](0,.35) -- (.25,0) -- (1,.7) -- (.25,.15) -- cycle;}

\definecolor{dkgreen}{rgb}{0,0.6,0}
\definecolor{gray}{rgb}{0.5,0.5,0.5}
\definecolor{mauve}{rgb}{0.58,0,0.82}

\SetAlFnt{\small}
\SetAlCapFnt{\small}
\SetAlCapNameFnt{\small}
\SetAlCapHSkip{0pt}
\IncMargin{-\parindent}

\acmDOI{0000001.0000001}

\begin{document}
\title{Constraint-Driven Modeling Enabling Dual Model Checking and Simulation for Discrete Event Systems}

\author{Soroosh Gholami}
\affiliation{%
  \institution{Arizona State University}
  \streetaddress{699 S Mill Ave}
  \city{Tempe}
  \state{AZ}
  \postcode{85282}
  \country{USA}}
\author{Hessam S. Sarjoughian}
\affiliation{%
  \institution{Arizona State University}
  \streetaddress{699 S Mill Ave}
  \city{Tempe}
  \state{AZ}
  \postcode{85282}
  \country{USA}}

\renewcommand\shortauthors{Gholami, S. and Sarjoughian H.S.}

\begin{abstract}
Verification and validation (V\&V) are crucial methods for evaluating the requirements and specifications of dynamical models that fulfill their intended purposes. Parallel Discrete EVent System Specification (PDEVS) is a system-theoretic modeling approach for creating modular, hierarchical component-based simulation models. In this paper, we introduce Constraint-DEVS, a method for creating bounded Parallel DEVS models that lend themselves, in addition to simulation, to model checking. We extend the DEVS-Suite framework to create Constraint-DEVS specifications which can then be model checked using a proposed state exploration protocol with the Parallel DEVS abstract simulator protocol. These capabilities, along with the support for non-determinism, complex data transfer, and performance-related property checking, make Constraint-DEVS and its accompanying DEVS-Suite a unique framework for the development, verification, and validation of discrete-event systems. In order to demonstrate this work, we developed and verified models of Network-on-Chip. Also, we detail behavioral design artifacts for the DEVS-Suite framework's hybrid model-checking and simulation engine.
\end{abstract}

%
%


%
%


\thanks{Authors' addresses: S. Gholami, Arizona Center for Integrative Modeling \& Simulation, Computer Science Department, School of Computing, Informatics, and Decision Systems Engineering, 699 S Mill Ave, Tempe, AZ 85281, US;
  H.S. Sarjoughian, Arizona Center for Integrative Modeling \& Simulation, Computer Science Department, School of Computing, Informatics, and Decision Systems Engineering, 699 S Mill Ave, Tempe, AZ 85281, US}

\maketitle

\section{Introduction}
\label{section:introduction}
It is generally necessary to develop Various models of a system during design. The dynamical models are evaluated using both verification and validation methods throughout incremental and iterative design stages. Designers use V\&V in order to achieve an acceptable, practical degree of assurance that the models accurately capture formulated requirements and specifications. Model validation is based on some reference model that could be mental (expectations of the designer from the model) or physical (the real system which is cloned into a model). However, for verification, one focuses on whether the model accurately reflects the specifications of the system. These methods are commonly stated as validation shows the right model is built, and verification shows the model is built right.

One can use various techniques for V\&V \cite{sargent2005verification,whitner1989guidelines} such as stress testing, Turing tests, and induction. A vast number of verification and validation methods are used in the community. They can be categorized as Informal, Static, Dynamic, Symbolic, Constraint, and Formal \cite{balci1994validation}. In this work, we focus on Simulation as the validation method and Model Checking as the verification method.

A simulation model is an executable representation of a real system. Some simulation models can be created using notional concepts while others may be grounded in mathematical formalisms such as DEVS \cite{zeigler2000theory} and Timed Automata \cite{alur2015principles}. While specialized simulators can simplify developing models and/or fast executions, they cannot be easily generalized for other uses. Models developed using mathematical formalisms, on the other hand, are very customizable and platform-independent. Simulation models may also be of different types with respect to the way they handle time, compositions, and control. Examples are hybrid, agent-based models, discrete-event discrete-time, and continuous-time approaches.

DEVS-Suite \cite{ACIMS-Software} is a discrete event simulation engine for parallel DEVS models. As a simulation engine, it has a variety of features such as the unique capability of generating superdense-time trajectories at run-time, animation, tracking (for monitoring and data collection), black-box unit testing, and debugging \cite{DEVS-Scripting}.

Formal model checking is a proof method to ensure a model satisfies certain properties. These models may represent critical systems responsible for important tasks. State explosion is a well-known problem when model checking is applied to large models \cite{burch1992symbolic}. Various methods are used to combat this problem and enable using model checking for time-sensitive and safety-critical systems to explore the entire state space \cite{alur2015principles}.

A common method in verifying formal models is to convert them to finite state machines. Modeling methods such as Timed Automata and Petri net \cite{james1981petri} correspond to a finite state machine. As mentioned, DEVS is not suitable for model checking because of the continuity of time and boundless state space. DEVS models are well-suited for simulation as carried out in tools such as DEVS-Suite. However, for model checking, a bounded state space is necessary.

Our contribution in this paper is the Constraint-DEVS model checking \cite{gholami2017modeling} and supported with the DEVS-Suite framework such that verification (in the form of model checking), in addition to validation (in the form of simulation), can be achieved.  The other contribution of this work is the creation of NoC (Network-on-Chip) models using Constraint-DEVS for both verification and validation. The third major contribution is the extension of the DEVS-Suite framework to support Constraint-DEVS modeling and verification via model checking. These enable as much as possible to seamlessly develop specifications supporting both evaluation methods (simulation and model checking) and realization within the DEVS-Suite framework.

\begin{figure}
    \centering
  	 \includegraphics[width=.5 \textwidth]{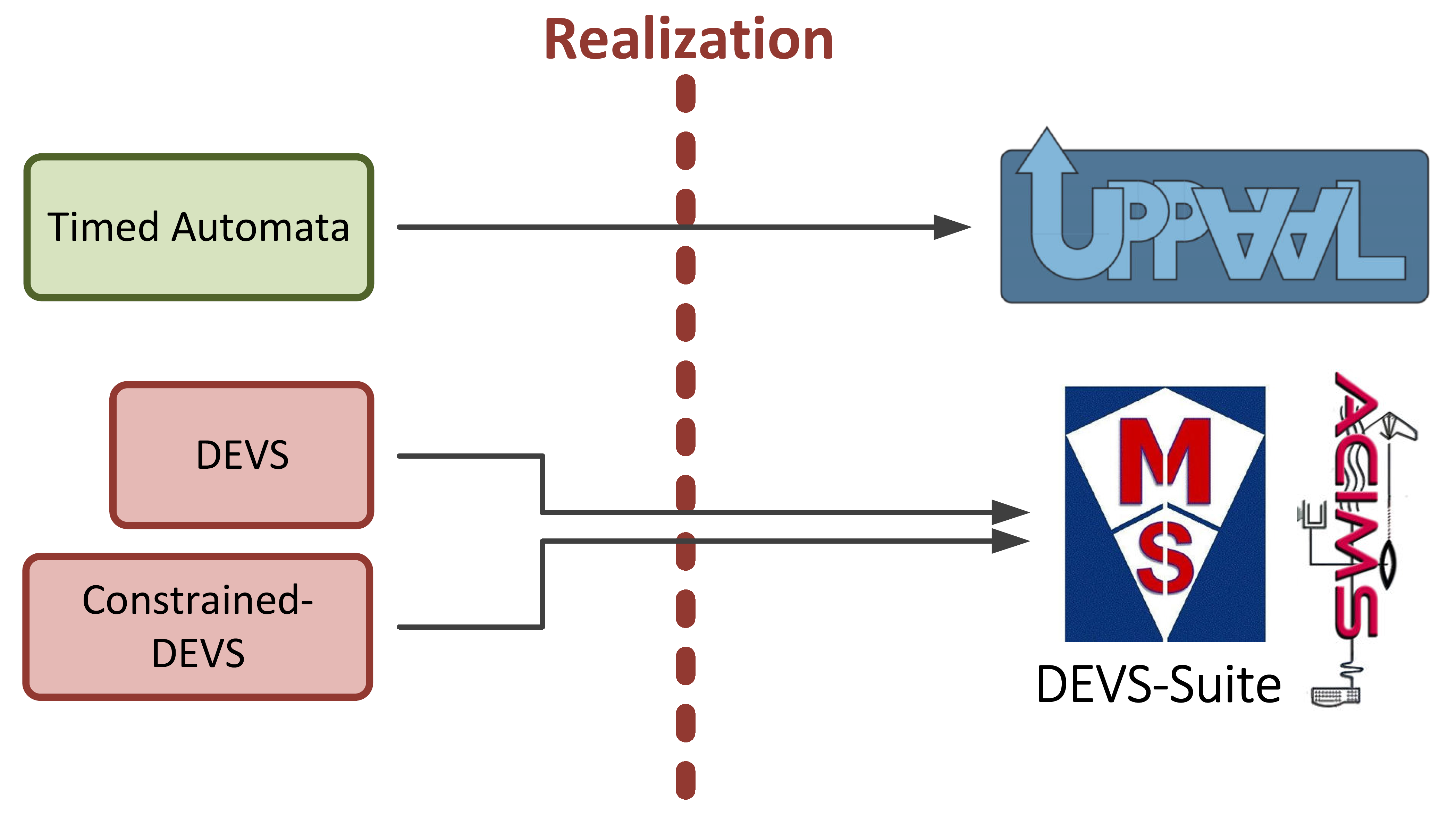}
    \caption{Realization of a version of Timed Automata in UPPAAL and DEVS/Constraint-DEVS in DEVS-Suite.} \label{fig:realizationOfDEVSTA}
\end{figure}

For the NoC models created via Constraint-DEVS, we define NoC-specific properties and verify them using DEVS-Suite. These models demonstrate the added capability of this modeling, simulation, and model checking environment. In order to provide the reader with a more comprehensive view, we develop a similar model in UPPAAL \cite{behrmann2014tutorial} using Timed Automata and compare these two environments with respect to their capabilities in modeling, simulation, and model checking of complex systems of the type considered here. As depicted in Figure \ref{fig:realizationOfDEVSTA}, it is important to note that each of UPPAAL and DEVS-Suite is a realization of Timed Automata and Constraint-DEVS, respectively. Other realizations exist but we limit this comparison only to these environments.

The remainder of this paper is organized as follows. In Section \ref{sec:background}, we provide the reader with some background on modeling, simulation, and model checking from the vantage point of the Parallel DEVS approach. In Section \ref{sec:RelatedWork}, the related work is discussed. We describe Constraint-DEVS as a verifiable variant of DEVS in Section \ref{sec:ConstrainedDEVS}. In Section \ref{sec:nocExample}, we detail models of NoC developed in UPPAAL and DEVS-Suite and contrast their modeling, simulation, and model checking capabilities. In Section \ref{sec:discussion}, we describe our contribution relative to the existing frameworks and practices for generic modeling, validation, and verification. Finally, in Section \ref{sec:conclusion}, we conclude this paper and discuss some future work. We detail the software design aspects of the extension for the DEVS-Suite framework (as a realization of the Constraint-DEVS specification and an engine for simulation and verification) in Appendix \ref{sec:verificationInDEVSSuite}.

\section{Background}
\label{sec:background}
In this section, we discuss how systems are specified using DEVS, how they are executed for simulation, and what is required for supporting model checking. An illustrative example of a circular buffer is provided to show the modeling concepts discussed in the remainder of this paper. Circular buffers are a common method of buffer implementation in hardware chips. We present this model from the library of Network-on-Chip (NoC) DEVS models \cite{gholami2016multi}.

\subsection{DEVS Modeling}
Discrete EVent System Specification (DEVS) is a hierarchical, continuous time formalism devised for modeling and simulation of reactive systems \cite{zeigler2000theory}. Systems can be modeled as a set of communicating automata in a hierarchical fashion using atomic and coupled DEVS models.
$$Atomic = \langle X^b, S, Y^b, \delta_{ext}, \delta_{int}, \delta_{conf}, \lambda, ta \rangle.$$

In this description, the input ports with events, output ports with events, and sequential state set are represented by $X^b$, $Y^b$, and $S$. External transition function is defined as $\delta_{\mathit{ext}}\,{:}\, Q \times X^b \to S$ where $Q=\{(s,e)|s \in S, 0 \leq e \leq ta(s)\}$. This function is a mapping between the occurrence of a bag of external events on one or more input ports and the sequential state set at any instance of time. The internal transition function, $\delta_{\mathit{int}}\,{:}\, S \to S$, defines how the model reacts to internal events. The confluent function, $\delta_{\mathit{conf}}\,{:}\, Q \times X \to S$, handles the occurrence of simultaneous (internal and external) events. The confluence function determines which event (internal or external) has priority over the other. Simultaneous external events must be handled by the external transition function. External events are presented in a set (without any order); it is the responsibility of the modeler to define order among events if necessary. The output function ($\lambda\,{:}\, S \to Y^b$) specifies output generation by mapping the state set to a bag of output events on one or more output ports at any instance of time. Finally, the time advance function, $ta\,{:}\, S \to \mathbb{R}_{0,\infty}^{+}$, specifies the timing behavior of the system. In the remainder of this paper, $X$ and $Y$ replace $X^b$ and $Y^b$ for brevity.

Hierarchical structures in DEVS are made possible through coupling input and output ports of atomic/coupled models subject to no direct feedback coupling (feedback from an output port to the input port of the same atomic/coupled model). Coupled models do not contain state information; they only specify how components are placed and communicate with one another under a strict hierarchical tree structure.

$$Coupled = \langle X, Y, D, \{M_d\}, EOC, EIC, IC \rangle.$$

The input and output for DEVS coupled models have identical specifications as those for atomic model. \textit{D} is an index set (component names) of the atomic/coupled models contained in the coupled model. $\{M_d\}$ is the set of internal atomic/coupled models for \textit{D}. Finally, the three sets \textit{EOC}, \textit{EIC}, and \textit{IC} specify a set of external output port couplings, a set of external input port couplings, and a set of internal couplings (for internal couplings between atomic/coupled models within $\{M_d\}$), respectively.

\begin{figure}
    \centering
  	 \includegraphics[width=.6 \textwidth]{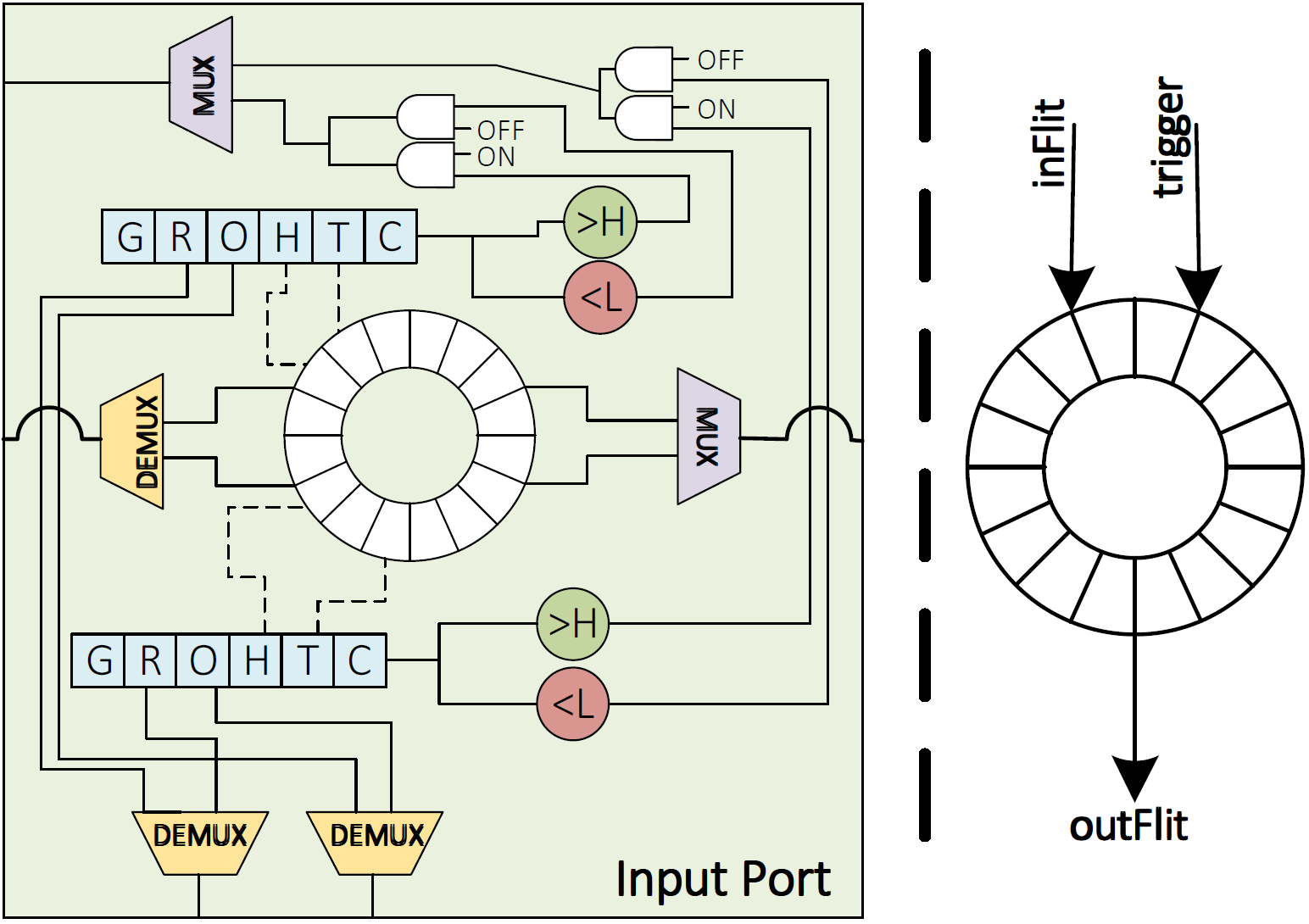}
    \caption{NoC switch input module utilizing a circular buffer (left) with its simplified diagram (right).} \label{fig:circularBuff}
\end{figure}

To see DEVS in action, we provide a model of a NoC circular buffer. It is one way of implementing the storage capability in an input module of a NoC router \cite{ni1998circular}. It is utilized to simplify the operation of an NoC with multiple virtual channels. These buffers hold flits (packets of data) waiting to be routed. Figure \ref{fig:circularBuff} shows the circular buffer in a NoC input module.

We model a simplified version of the circular buffer with two input ports (i.e., one for storing input data (flit) and another for triggering sending out stored flits) and one output port for outputting flits. The buffer operates in FIFO mode (see Listing \ref{circularBuff-DEVSModel}). Note that the modeling presented in Listing \ref{circularBuff-DEVSModel} is a DEVS model and is only appropriate for simulation. Later in this paper, we present a model of a circular buffer in Constraint-DEVS and its realization in DEVS-Suite suitable for both model checking and simulation.

The state set has variables for the head index, tail index, and a data structure for holding the data. In this model, the receipt of a new flit invokes the external transition function. This transition results in storing the flit and increasing the tail index. A trigger event invokes and ultimately causes outputting the head of the queue. Obviously, inserting flits to a full buffer and retrieving flits from an empty one is not allowed. We set the value of the time advance function (\textit{ta}) to $0.5$ in state \textit{Active}, which means, upon receiving the \textit{trigger} signal, it takes $0.5$ cycle for the buffer to output the head flit.

This model does not contain hierarchy; it is an atomic model. We chose to exclude hierarchy from the example since this would be sufficient to show the model checking protocol with respect to the simulation protocol.

\begin{lstlisting}[mathescape=true,label=circularBuff-DEVSModel,caption=Model of a single channel circular buffer in DEVS.]
    $S=\overbrace{\{\mathit{Active, Idle}\}}^\text{phase} \times \overbrace{\sigma}^\text{sigma} \times \overbrace{\{0,1\}^{*}}^\text{flitBuffer} \times \overbrace{\mathbb{N}}^\text{head} \times \overbrace{\mathbb{N}}^\text{tail} \times \overbrace{\beta}^\text{outFlit}$
    $X=\big\{\big(\mathit{inFlit}, \{0,1\}^{*}\big), \big(\mathit{trigger}, 1\big)\big\}$
    $Y= \big\{(\mathit{outFlit}, \{0,1\}^{*})\big\}$
    $\delta_{\mathit{ext}}\big((\mathit{Idle}, \sigma, \mathit{flitBuffer}, \mathit{head}, \mathit{tail}, \varnothing), e, (\mathit{inFlit}, x)\big) =$
        $\begin{cases}
 		     (\mathit{Idle}, \infty, \mathit{flitBuffer}.x, \mathit{head}, \mathit{tail}+1, \varnothing)  \hspace{6mm} \textbf{if} \; \mathit{head} \neq \mathit{tail}  \\
		     (\mathit{Idle}, \infty, \mathit{flitBuffer}, \mathit{head}, \mathit{tail}, \varnothing)  \hspace{6mm} \textbf{if} \; \mathit{FULL}
        \end{cases}$
    $\delta_{\mathit{ext}}\big((\mathit{Idle}, \sigma, \mathit{flitBuffer}, \mathit{head}, \mathit{tail}, \varnothing), e, (\mathit{trigger}, x)\big) =$
        $\begin{cases}
 		     (\mathit{Active}, .5, \mathit{flitBuffer}, \mathit{head}+1, \mathit{tail}, \mathit{flitBuffer}.\mathit{head})  \hspace{6mm} \textbf{if} \; \mathit{head} \neq \mathit{tail}  \\
		     (\mathit{Idle}, \infty, \mathit{flitBuffer}, \mathit{head}, \mathit{tail}, \varnothing)  \hspace{6mm} \textbf{if} \; \mathit{EMPTY}
        \end{cases}$
    $\delta_{\mathit{conf}}\big((\mathit{phase}, \sigma, \mathit{flitBuffer}, \mathit{head}, \mathit{tail}, \beta), e, (\mathit{PORT}, x)\big) =
       \delta_{\mathit{ext}}\big(\delta_{\mathit{int}},0,(\mathit{PORT},x)\big) $
    $\delta_{\mathit{int}}\big((\mathit{phase}, \sigma, \mathit{flitBuffer}, \mathit{head}, \mathit{tail}, \beta) = (\mathit{idle}, \infty, \mathit{flitBuffer}, \mathit{head}, \mathit{tail}, \varnothing)$
    $\lambda(\mathit{Active}, \sigma, \mathit{flitBuffer}, \mathit{head}, \mathit{tail}, \beta)=(\mathit{outFlit}, \beta)$
    $\mathit{ta}(\mathit{Active}, \sigma, \mathit{flitBuffer}, \mathit{head}, \mathit{tail}, \beta)=0.5$
\end{lstlisting}

\subsection{DEVS Simulation}
\label{sec:devsSimulation}
The simulation protocol is responsible for executing DEVS atomic models. The reason we included this section in this paper is that we incorporate the DEVS-Suite simulation engine within the model checking engine. Our state exploration protocol wraps around the simulation protocol and uses it for cycle-by-cycle execution. The mechanism will be elaborated later in this paper.

DEVS-Suite mechanism for DEVS execution relies on simulator and coordinator protocols. The execution of every atomic model is handled by its dedicated simulator; similarly, every coupled model has its dedicated coordinator.

\begin{figure}
    \centering
  	 \includegraphics[width=.55 \textwidth]{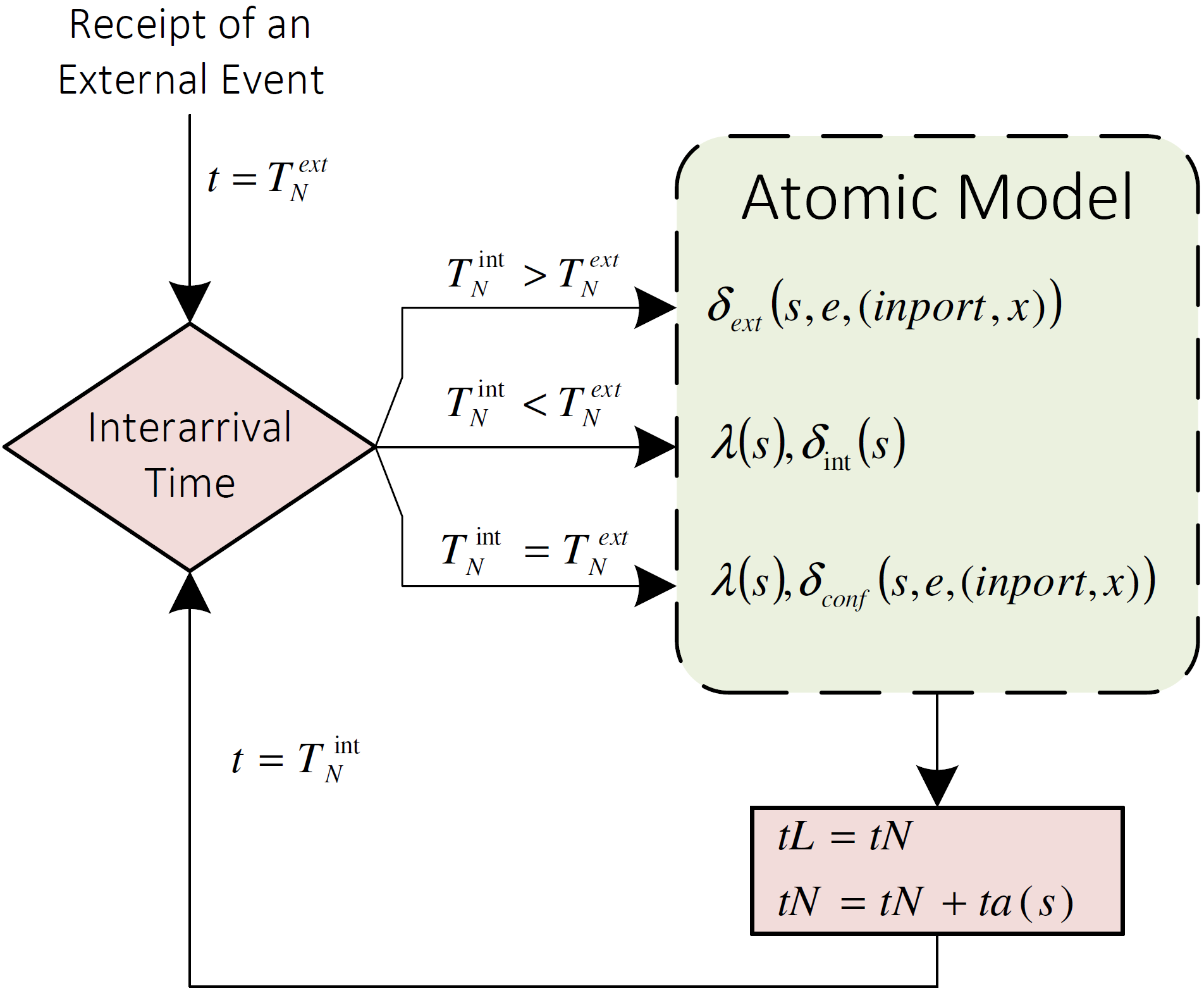}
    \caption{Time management in the coordinator module (with solid borders and in red) and invocation of the atomic model functions (with dashed borders and in green) in the simulator module.} \label{fig:DevsSimulationProtocol}
\end{figure}

Simulators and coordinators manage timing and choice of functions to be executed, as well as sending and receiving input and output events for atomic and coupled models they supervise, under the control of a root-coordinator assigned to the highest-level coupled model. The behavior specific to every atomic model, specified as functions, is always invisible to the simulator protocol. The control of the simulator on the timing aspect of an atomic model is depicted in Figure \ref{fig:DevsSimulationProtocol}. Time of last event (\textit{tL}), time of next event (\textit{tN}), and external input events are used by the simulator to control the execution of an atomic model. Upon receipt of an event (the one with the earliest scheduled time, i.e., \textit{tN}) by an atomic model, the appropriate function (external transition, internal transition, or confluence) is determined and executed. As shown in Figure \ref{fig:DevsSimulationProtocol}, this decision is made by comparing the time instance of the event with the expected time of the next event.  The output function is invoked to generate outputs, if there are any, before the execution of the internal transition function. After executing the function(s), the time of the next event (\textit{tN}) is set and the model moves on to the next cycle of the simulation.

This is the simulation protocol used in DEVS-Suite for DEVS model execution. However, this protocol is not appropriate for model checking. In Section \ref{sec:ConstrainedDEVS}, we describe how we incorporate this simulation protocol into the model checking protocol.

A sample simulation scenario for the model of circular buffer (presented earlier) is shown in Table \ref{table:circularBufferSimulation}. In this scenario, a number of random external events are injected into the model. Some of the state variables and the output of the model are included in the table to illustrate how the execution protocol simulates an atomic model.

\begin{table}
\centering
\caption{A sample simulation scenario for circular buffer.}
\label{table:circularBufferSimulation}
    \begin{tabular}{| c | c | c | c | c | c | c |}
    \hline
 \rotatebox[origin=c]{90}{\parbox[c]{1.5cm}{\centering Time}} & \rotatebox[origin=c]{90}{\parbox[c]{1.5cm}{\centering inFlit}} & \rotatebox[origin=c]{90}{\parbox[c]{1.5cm}{\centering Trigger}} & \rotatebox[origin=c]{90}{\parbox[c]{1.5cm}{\centering Buffer Phase}} & \rotatebox[origin=c]{90}{\parbox[c]{1.5cm}{\centering Buffer Output}} & \rotatebox[origin=c]{90}{\parbox[c]{1.5cm}{\centering Head Index}} & \rotatebox[origin=c]{90}{\parbox[c]{1.5cm}{\centering Tail Index}}\\
   	\hline
   \hline
 0 & - & - & \textit{Idle} & - & 0 & 0 \\[1mm]
 2 & A & - & \textit{Idle} & - & 0 & 1 \\[1mm]
 3 & B & - & \textit{Idle} & - & 0 & 2 \\[1mm]
 3 & C & - & \textit{Idle} & - & 0 & 3 \\[1mm]
 4 & D & 1 & \textit{Active} & - & 0 & 4 \\[1mm]
 4.5 & - & - & \textit{Idle} & A & 1 & 4 \\[1mm]
 6 & E & 1 & \textit{Active} & - & 1 & 5 \\[1mm]
 6.5 & - & 1 & \textit{Active} & B & 2 & 5 \\[1mm]
 7 & F & 1 & \textit{Active} & C & 3 & 6 \\[1mm]
 7.5 & - & - & \textit{Idle} & D & 4 & 6 \\[1mm]
	\hline
  \end{tabular}
\end{table}

\subsection{Requirements for DEVS Model Checking}
\label{sec:requirements}
In the previous sub-sections, we illustrated how systems are modeled with DEVS specification and simulated via the DEVS simulation protocol. The question we answer in this paper is can one use the DEVS specification and the simulation protocol (as presented earlier) for model checking? But before answering the question, we first present the requirements that are needed for model checking. We also highlight the expressiveness of the DEVS formalism and empower DEVS-Suite to support Constraint-DEVS modeling and verification.

Considering any DEVS model's state space as a graph with transitions as arcs and states as nodes, the entire graph must be explored. Therefore, we need to ensure that the model possesses a finite state space. This requires the number of internal/external transitions (as arcs) and states (as nodes) to be finite. For the finite internal transitions requirement, the time advance function for every atomic model must be defined relative to a discrete time base. Given $|S|<\infty$ for $ta:S\to R'$, where $R'$ is a finite set (i.e., $R'\in \mathbb{R}$), there can exist a finite number of internal transitions. For the external transitions, DEVS allows input events to be received at any arbitrary instance of time. Therefore, external events are constrained to arrive only at discrete time instances. Also, the values for any input port must be bounded. As for the last condition, we use bounded state variables to represent the state. These three constraints are discussed in more detail in Section \ref{sec:ConstrainedDEVS}.

DEVS-Suite as a simulation engine for DEVS models cannot enforce the above three constraints. Therefore, the changes we propose (and discuss in detail in Section \ref{sec:ConstrainedDEVS}) are required for DEVS-Suite to support model checking. We wish our tool (for modeling, simulation, and model checking) to support \textit{non-determinism}, \textit{stochasticity}, \textit{complex data transfer} (information flow) and \textit{property checking} beyond those supported in LTL (Linear Time Interval), CTL (Computation Tree Logic), or CTL* (Linear and Temporal Control Logic) \cite{TemporalLogic}. The restrictions of other approaches (DEVS-based and non-DEVS-based formalisms) for simulation and model checking are discussed in Section \ref{sec:RelatedWork}.

\section{Related Work}
\label{sec:RelatedWork}
Among the most widely used modeling methods used for model checking are Petri nets and Timed Automata. Petri nets models are constructed with places, arcs, and transitions. Similarly, basic Time Automata models are made up of states, clocks, transitions, and actions. Various environments such as Alpha/Sim \cite{moore1995alpha} for Petri nets and UPPAAL \cite{behrmann2014tutorial} for Time Automata, support modeling, simulation, and model checking for these methods. In both tools, formal languages (CTL for Alpha/Sim and CTL* \cite{alur1993model} for UPPAAL) are used for property expressions. A limitation of Petri nets and Time Automata is their inability to handle complex data types \cite{lanotte2005timed}. Also, these two methods (and their respective tools) do not support modeling the structure of a given system as they only model the behaviors of concurrent systems. These two major shortcomings are addressed together with the Constraint-DEVS and DEVS-Suite.

While LTL, CTL, or CTL* are all strong languages with succinct syntax and semantics for property expression, they are not expressive enough to support complex QoS properties such as the average latency of packets transmitted from a specific node in a network. The reason is that unlike state-based properties (which evaluate the state of the model at a single instance of time), QoS properties require the verifier to monitor the model over a period of time; this is not supported by variants of linear or computation tree logics \cite{holcomb2014compositional}. We proposed using the experimental frame (EF) \cite{rozenblit1991experimental} in order to address the expressiveness problem \cite{gholami2017modeling}.

The concept of experimental frame (EF) is used in simulation frameworks for validation. We are incorporating it for verification purposes as well as validation in DEVS-Suite. However, if the EF has bugs, the entire evaluation is compromised.  Therefore, EF as a formal language is considered for verifying DEVS model \cite{foures2013simulation}. This approach can be used to ensure the correctness of the EF itself.

As for DEVS variations suitable for model checking, FD-DEVS \cite{hwang2009reachability}, RTA-DEVS \cite{saadawi2013principles}, and FPDEVS \cite{seo2015integrating} approaches have been proposed. These incorporate some kind of model conversion in order to make model checking possible. FD-DEVS models are converted to PROMELA \cite{PROMELA-language} and then verified using SPIN \cite{holzmann1997model} and LTL property language. RTA-DEVS models are converted to Time Automata and verified in UPPAAL. The Finite Probabilistic DEVS (FPDEVS) is a variation of FD-DEVS in which the choice of next state is made probabilistically. In FD-DEVS, the choice of next state is deterministic through a lookup table. In FPDEVS, an atomic model determines the next state by creating a cumulative distribution function from each transition's probability value and then choosing one based on the value of a random number generator. While these DEVS-based approaches do not have the two major shortcomings of Petri nets and Time Automata, the need for conversion eventually brings those problems to the surface. The models that are in the end verified still lack support for complex data and structural modeling. In addition, these DEVS-based approaches still rely on tree logic for property expression which, as mentioned before, has its limitations in terms of expressing complex properties.

Another method of approaching the cycle-accurate simulation of NoC has been explored in \cite{8943201} via the Cellular Automata (CA) modeling framework. Similar to this work, NoC components are formally modeled and a modified simulation algorithm is devised to execute the model. Performance measures can be defined and extracted from the simulation. Comparing this approach to RTL-level NoC modeling and simulation, the authors claim significant execution time improvements and ease of modeling. While this work has commonalities with our approach with respect to formal modeling of NoC components and modified execution algorithms which both ensure efficiency and cycle-accurate simulations, there are fundamental differences between the two. One of the primary goals of this work is to enable model verification using state exploration while keeping the simulation functionality intact. While in \cite{8943201}, the main goal has been to create a fast modeling/simulation framework using CA requires models to be at coarse-grain and validated using fine-grain models.

The problem of state explosion is well-known in model checking. Large hybrid systems are impractical to be verified using model checking. That is why verification approaches based on simulation \cite{kapinski2015simulation} and falsification \cite{plaku2009hybrid} are introduced.
A category of tools provides visualization and analysis of simulation traces. An example is Traviando \cite{kemper2006traviando}. However, the purpose here is to develop a tool that supports simulation and model checking of dynamic models. Therefore, trace-based and debugging tools such as Traviando are considered out of scope.

\section{Constraint-DEVS}
\label{sec:ConstrainedDEVS}
Model verification in DEVS entails four additional capabilities (relative to those already existing) within the DEVS framework: 1) bounded state configuration, 2) bounded input port configuration, 3) a finite number of internal/external events, and 4) state exploration protocol. Here we briefly discuss these four features. For additional details, we refer the reader to our earlier work \cite{gholami2017modeling}.

\subsection{State Variable Configuration}
State variables can be \emph{Primitive} and \emph{Compound}. A primitive state variable may only be of certain data types including Character, String, Double, Enumeration, Integer, and Boolean. Compound states (such as Hash Maps) are made by combining primitive and/or compound states. We suggested using \emph{regular expressions} to formulate compound state variables using a set of primitive ones acting as the alphabet for this language. For example, a queue of size 8 that holds strings (each with 16 characters) can be specified as $\mathit({Char}^{16})^8$. Products of primitive state variables can make compound ones.

The benefit of such a formulation (regular expressions) is that the state exploration protocol can calculate the size of the state space and iterate through all combinations. The size of the state space is also impacted by port values and time. Thus, restrictions must be put on those as well as the state exploration protocol must explore all possible combinations of them.

\subsection{Port Configuration}
\label{subsec:PortConfig}
We use a similar technique to specify port types and value sets. It must be noted that we only need to specify the external input ports (used for events coming from outside of the model); all internal input and output ports are left unchanged since they are driven by internal atomic/coupled models and not directly affected by the state exploration protocol.

Ports can transfer primitive or compound data types to the models under inspection. Similar to state variables, we use regular expressions for specifying value sets for these ports. However, an additional NULL ($\varnothing$) value is always added to the value set. In any cycle in which the input port carries no data, $\varnothing$ is set as its value. As an example, an input port that carries strings (of size 5), can be specified as: $(\mathit{Char})^5 \cup \varnothing$.

The verification engine is again capable of calculating all possible port values. Any of these events can be applied to any given state of the model. The verification engine applies all possible combinations of inputs to all reachable combinations of state variables for a reachability analysis.

\subsection{Finite Number of Internal/External Events}
\label{subsec:FiniteInternalExternal}
As explained in Section \ref{sec:requirements}, the number of internal transitions can be made finite by restricting the use of continuous time-base for Time Advance (\textit{ta}) function (see Section~\ref{sec:background}.

As for the external events, the receipt of external input events should occur at discrete time instances to limit the number of transitions from each state. Similar to internal transitions, the use of a continuous variable for elapsed time (\textit{e}) can result in having external output events at arbitrary time instances (i.e., an infinite number of output events). The continuity of time for atomic/coupled DEVS models stays the same. As long as there are no converging sequences of time-advance function values, the state space of the model stays finite.

\begin{figure}
    \centering
  	 \includegraphics[width=.6 \textwidth]{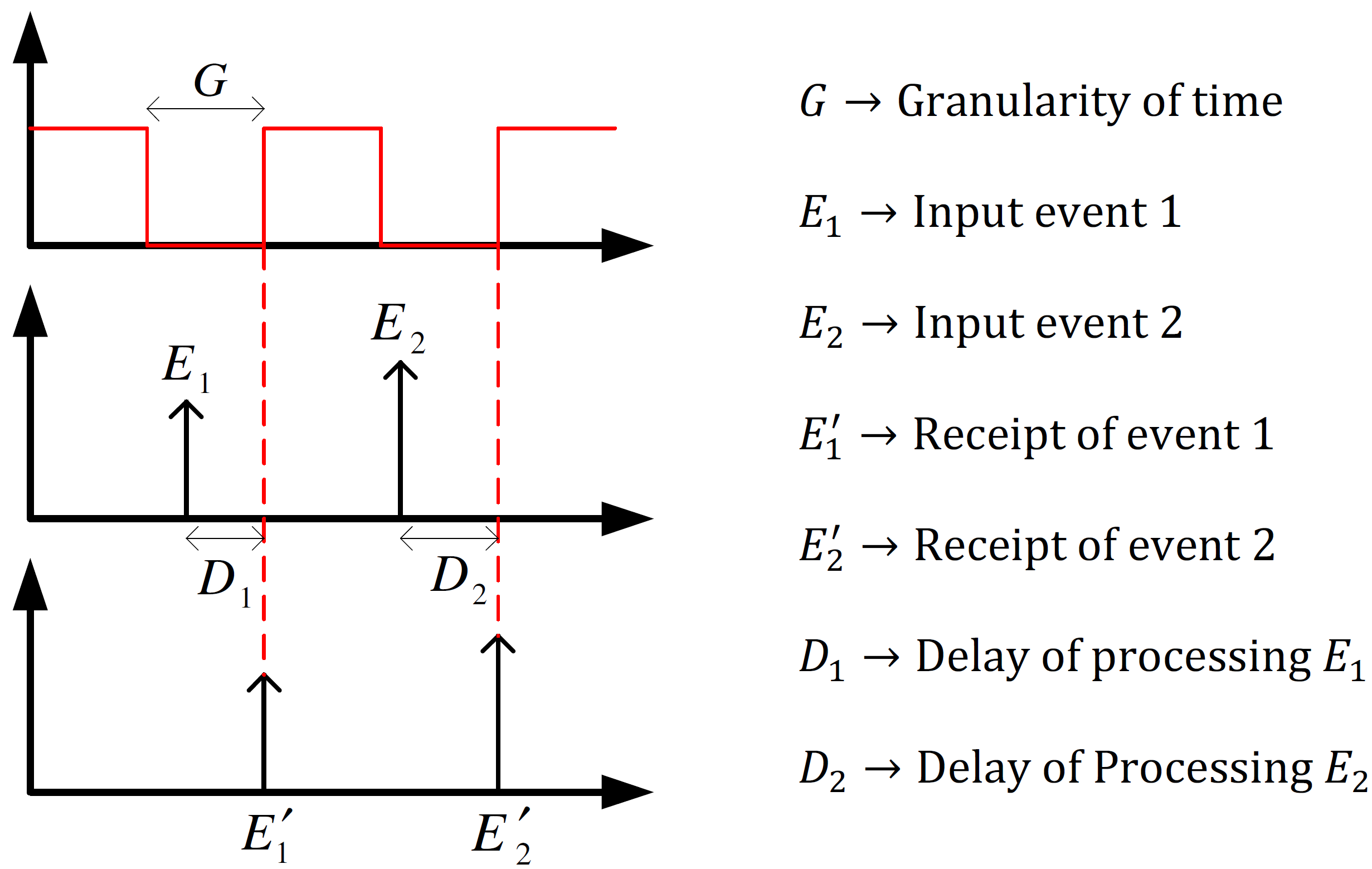}
    \caption{Discrete time external input event processing.} \label{fig:discreteTimeEventProcessing}
\end{figure}

Discretizing the time of external inputs requires delaying event occurrences before they are received and processed. As an example, in Figure \ref{fig:discreteTimeEventProcessing}, if $E_1$ occurs at relative time instance $G-D_1$ and $G-D_1 \neq 0$, then $E_1$ is changed to ${E}^\prime _1$ such that its time instance is the next integer multiplier of the model's temporal resolution $G$.

\subsection{State Exploration Protocol}
The state exploration algorithm repeats a set of tasks to cover the entire state space of the model. There are three state sets: unvisited ($Q$), visited ($V$), and unsafe ($U$). The state exploration starts with a set of initial states stored in $Q$ and repeats certain steps until $Q$ is empty. During this process, if any undesirable state is visited (stored in $U$), the process is terminated and the user is notified. Although the state exploration protocol terminates after observing the first undesirable state, it can be reconfigured to continue to explore the global state set. Upon termination, this alteration of the protocol provides a set of invalid states visited in the course of exploration.

As explained earlier in Sections \ref{subsec:PortConfig} and \ref{subsec:FiniteInternalExternal}, internal events are finite (and are defined using regular expressions) and time is discretized. The state exploration protocol utilizes the finiteness of time and input events to apply all possible inputs to the model during model-checking.

Listing \ref{state-exploration-protocol} presents the state exploration protocol for Constraint-DEVS models \cite{gholami2017modeling}. MOD is the target model. At each cycle of execution, a state is taken from $Q$ (step 4) and all possible inputs are applied to it (steps 5-16). After this is done, the state is moved to the set of visited state set (step 17) and a new cycle is started.

Earlier in this paper (Section \ref{sec:devsSimulation}), we mentioned that the state exploration protocol wraps around the simulation protocol and uses it for cycle-by-cycle model execution. This happens at step 8 of Listing \ref{state-exploration-protocol}. During this step, the model is simulated for only \underline{1 cycle} (with the inputs provided and the state which is set). Then the simulation is stopped and the state exploration protocol takes over again. So, simulation is used within model checking; however, this is a controlled execution. The model is executed only for 1 cycle after the state is set and inputs are injected.

\begin{lstlisting}[mathescape=true,label=state-exploration-protocol,caption= State exploration protocol for Constraint-DEVS models.]
$\textbf{Input:}\; \text{MOD:} \mathit{Verifiable}, \text{GEN:} \mathit{VerifierGen} $
$\textbf{Output:}\; \mathit{invalidState}:\;\mathit{StateVar}$
$\textbf{Initialization:}\; \mathit{instantiate} \; \textit{Q}, \; \textit{V}, \;$ and $\textit{U}\; ; \; \mathit{invalidState} \gets null$

1.	$\textbf{add} \; \text{MOD}.\mathit{initialStates}$ to $\textit{Q}$
2.	$\textbf{add} \; \text{MOD}.\mathit{unsafeStates}$ to $\textit{U}$
3.	$\textbf{while} \; \textit{Q} \neq \varnothing \; \textbf{do}$
4.	      $\textit{state-event} \gets Q.\mathit{head} ( )$
5.	      $\textbf{while} \; \textit{state-event}.\mathit{inputSet} \neq \varnothing \; \textbf{do}$
6.	             $\text{MOD}.state \gets \textit{state-event}.\mathit{state}$
7.	             $\text{GEN}.output \gets \textit{state-event}.\mathit{inputSet}.\mathit{head} (\;)$
8.	             $\textbf{call} \; \mathit{simulate} (\;)$
9.	             $\textbf{if} \; \text{MOD}.\mathit{state} \in \textit{U} \;\textbf{then}$
10.	                     $\mathit{invalidState} \gets \text{MOD}.\mathit{state}$
11.	                     $\textbf{return} \; \mathit{invalidState}$
12.	             $\textbf{end if}$
13.	             $\textbf{if} \; \text{MOD}.\mathit{state} \notin \textit{Q} \; \land \; \text{MOD}.\mathit{state} \notin \textit{V} \; \textbf{then}$
14.	                     $\textbf{add} \; \text{MOD}.\mathit{state}$ to $\textit{Q}$
15.	             $\textbf{end if}$
16.	      $\textbf{end while}$
17.	      $\textbf{add} \; \textit{state-event}$ to $\textit{V}$
18.	$\textbf{end while}$
19. $\textbf{return} \; \mathit{invalidState}$
\end{lstlisting}

\subsubsection{Constraint-DEVS for Coupled Models}
As mentioned earlier, Constraint-DEVS can be used to model both atomic and coupled systems. Moreover, we can use the simulation engine and the state exploration protocol (discussed in Section \ref{sec:selectiveStateExploration}) to validate and verify a model using simulation and model checking. The state of a coupled model is defined to be set of all of its internal models (atomic or coupled). All internal couplings are ignored by the state exploration protocol as they are driven by atomic/couple models. State exploration follows the protocol provided in Listing \ref{state-exploration-protocol} by treating any model as an atomic model. The algorithm goes through the entire set of possible states and applies all possible inputs to the input ports.

\subsubsection{Selective State Exploration}
\label{sec:selectiveStateExploration}
One important aspect of modeling with Constraint-DEVS in DEVS-Suite is the flexibility it offers in the specification of state, input, and output. Model state variables are marked as \emph{explorable} or \emph{unexplorable} (at initialization). An \emph{explorable} state variable is considered as a part of the state of the model, and therefore it is explored and adds to the collective state space size. An \emph{unexplorable} state variable operates similarly to the explorable one except that it is invisible to the outside world.

Within the model, both the explorable and unexplorable state variables are treated in the same way. Their values are tracked, changed, and used to make decisions. However, outside the model (particularly from the viewpoint of the verification engine), a change in the value of an explorable state variable is considered a new state while the change in the value of the unexplorable one is ignored.

\begin{figure}
    \centering
  	 \includegraphics[width=.7 \textwidth]{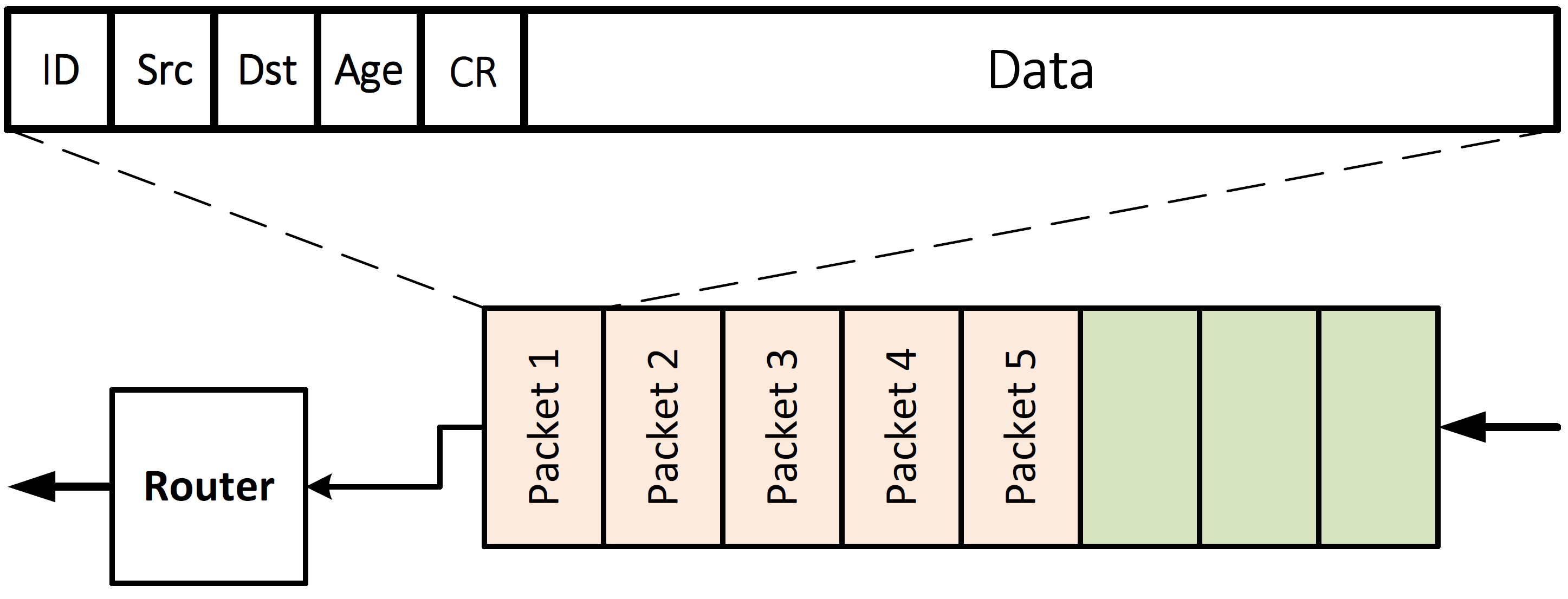}
    \caption{Waiting queue of a switch and the structure of packets it stores.} \label{fig:switchQueueAndStateFlexibility}
\end{figure}

Depending on whether or not a state variable needs to be considered in the state space, it is marked as explorable or unexplorable. As an example, the queue of a switch component in NoC holds a number of packets waiting to be routed. This is depicted in Figure \ref{fig:switchQueueAndStateFlexibility}. As illustrated, each packet contains an ID (unique), source node, destination node, age, computation requirement, and data. Now, the state of a queue with capacity 8 is the possible combinations of data it can hold. This will make the number of states for this queue an immensely large number. For a network size of 6, 100 packet IDs, 5 values for RC, 3 values for age, and 2 bytes of data, the queue's state space will be $100 \times 6^8 \times 6^8 \times 5 \times 3 \times 2^{16}$.

However, a switch does not care about the ID, source node, age, computation requirement, and the data that the packet carries. The only notable information for the switch is the destination node using which it routes the packet out. Therefore, the state space of the queue can be reduced to the combination of destination values it can hold. For example, for a network of 6 nodes, the state space of the queue can be rewritten as $6^8 \approx 1M$.

DEVS-Suite provides a simple way to change the role of each state variable at initialization. Therefore, the state space of the model can be easily managed for larger and more complex systems. Changing the role of a state variable does not impact the behavior of the model and its validation via simulation. This is a significant capability compared with other approaches such as Timed Automata and Petri nets. In those modeling methods, one cannot change the size of the state space of a model other than by creating a new model with fewer states.

\subsection{Trace Analysis}
Using the protocol presented in Listing \ref{state-exploration-protocol}, the verification engine explores the entire reachable portion of the state space (starting from a set of initial states). During the exploration process, a transducer constructs the reachability graph. In this graph, nodes (set \textit{V}) contain state information and edges (set \textit{E}) are transitions. After the exploration is completed, the transducer may analyze the graph (using graph algorithms), draw conclusions, and verify properties.

\begin{figure}
    \centering
  	 \includegraphics[width=.7 \textwidth]{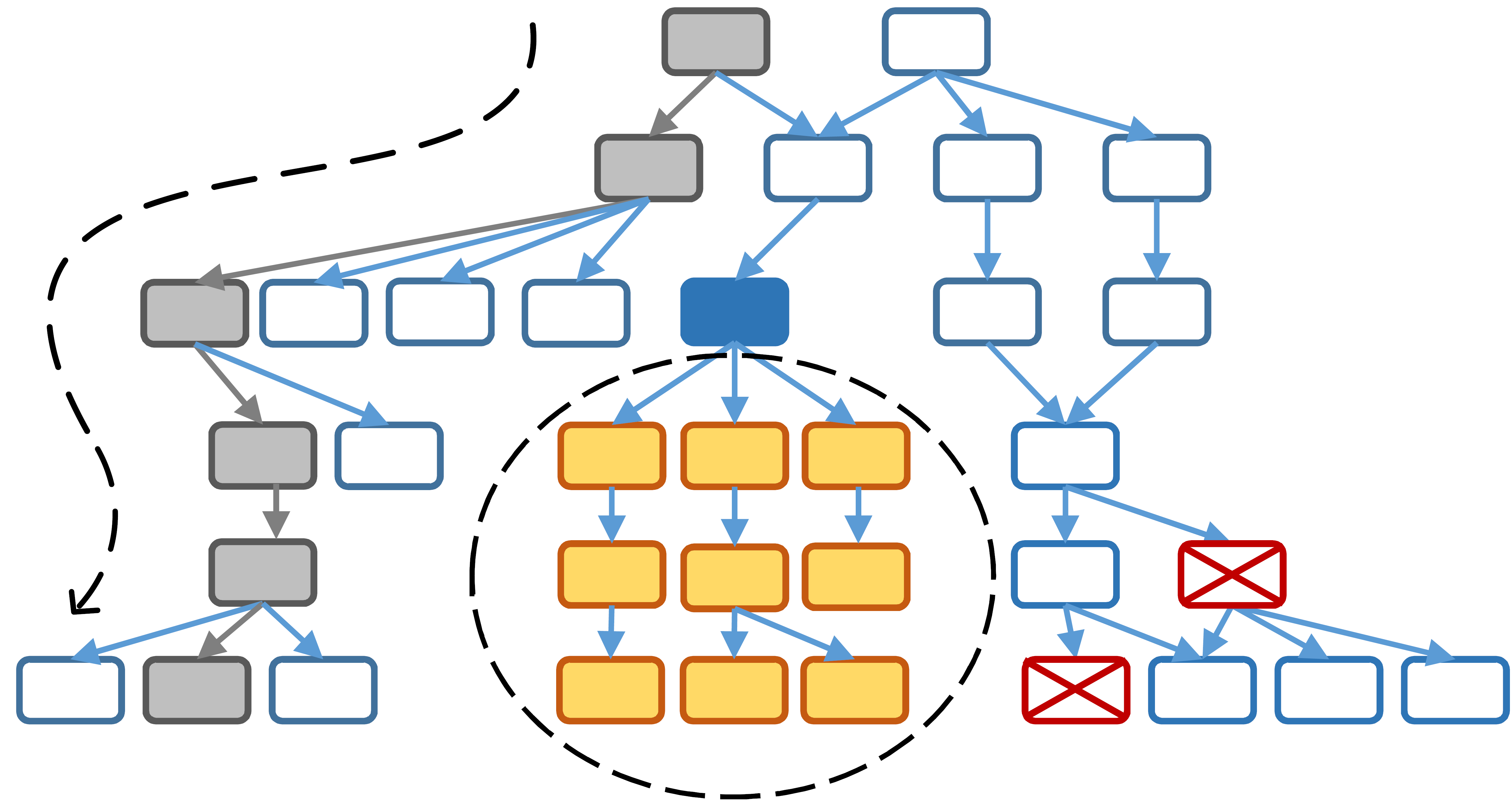}
    \caption{Trace analysis on a pruned reachability graph.} \label{fig:traceTreeAnalysis}
\end{figure}

By converting the graph to a set of traces (by pruning cycles), the transducer may verify properties related to individual states, paths, or subtrees. Figure \ref{fig:traceTreeAnalysis} depicts a set of traces resulting from two initial states. The transducer verifies state-based properties by looking at individual states (nodes marked X), path-based properties by looking at one or more paths (dashed arrow), and tree-based ones by looking at a subtree (dashed circle).

In Section \ref{sec:nocExample}, we conduct state-based (unsafe/invalid states) and path-based (overall component utilization) property checking for models of Network-on-Chip.

\subsection{Additional Features}
For state exploration, various techniques should be incorporated to reduce the size of the state space. Data exclusion is one of them. If data does not play a part in model verification, it can be removed to reduce the state space size. Data exclusion is done using the selective state exploration method explained in Section \ref{sec:selectiveStateExploration}.

For property expressions, we intended to introduce an approach that does not share the limitations of common languages such as LTL or CTL. We reuse the concept of Experimental Frame (EF) to create a generator and transducer pair in charge of exploring the state space and collecting the trace, respectively. The EF is implemented as non-DEVS components in Java.

Support for Constraint-DEVS and all the features mentioned in this section are realized in DEVS-Suite. For more on the software design aspects of the extension introduced into the DEVS-Suite framework, we refer the reader to Appendix \ref{sec:verificationInDEVSSuite}.

\section{Circular Buffer Example}
\label{sec:circularBuffExample}
As an example, we modeled the circular buffer (see Section \ref{sec:background}) with Constraint-DEVS and developed it in DEVS-Suite. This model is simple and is intended to highlight how the modeling, development, and verification phases work. The specification does not change much. Only restrictions are placed on state and input port values. In the case of a circular buffer, we assume that the size of the buffer is 8. Therefore, $\mathit{head} \in {0,1,2,...,7}$, $\mathit{tail} \in {0,1,2,...,7}$, and  $\mathit{flitBuffer} \in (\{0,1\}^{24} )^8$ (flits are assumed to be 24 bits). We also added a state variable called $\mathit{bufferStatus} \in {\text{full},\;\text{empty},\;\text{normal}}$. As for ports, the port \emph{trigger} can only send one type of signal and \emph{inFlit} can send flits to a number of different destinations.

For implementing the circular buffer in DEVS-Suite, three state variables \emph{head}, \emph{tail}, and \emph{bufferStatus} and two ports \emph{trigger} and \emph{inFlit} are modeled. Based on the number of possible inputs and the size of state variables (assuming a 100 node network with 99 destinations), the total number of state-events (combinations of states and input events which correspond to all states and the transitions among them) is $100\times2\times8\times8=12600$. The state space exploration algorithm presented in Listing \ref{state-exploration-protocol} iterates through all these states and gathers trace information.

\begin{figure}
    \centering
  	 \includegraphics[width=.65 \textwidth]{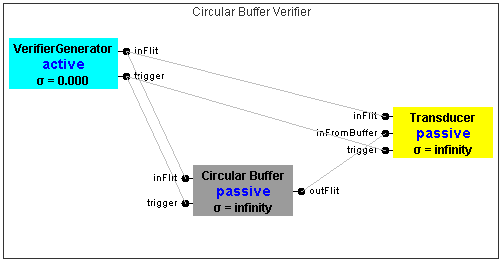}
    \caption{Circular buffer model verification.} \label{fig:circularBuffDEVSSuite}
\end{figure}

The generator and the verification engine are automatically generated by DEVS-Suite. The coupled component view of the verification model is shown in Figure \ref{fig:circularBuffDEVSSuite}. The transducer records incoming data, the state of the circular buffer, and the outputs generated by it for further analysis.

After the model of the circular buffer is implemented, we have the option of developing a custom transducer as well. The transducer monitors the output of the circular buffer to ensure outputs are generated at the right time (takes 0.5 cycle from the instance it is triggered), simultaneous events are handled properly (two events on two ports), correct setting of full/empty statuses, and correct updating of head/tail indexes. If developed correctly, the transducer is capable of finding model errors. As an example, for a basic circular buffer model, we had a small error in simultaneous event handling which could have easily gone unnoticed had it not been verified. The possibility of both simulating and model checking DEVS models in an integrated environment makes model development and evaluation simpler.

We ran a few instances of this verification with different parameters. Various scenarios can be created by changing configurations of the model under test (e.g., size of the buffer, types of incoming packets, and injection rate, ejection frequency).  This is in part due to maintaining the modularity of DEVS-Suite while extending it to support model checking. In each of these cases, all states are explored and the correctness of functionality is ensured. A few of these instances are shown in Table \ref{table:circularBufferVerification}.

The execution time for the state exploration linearly increases with the state space size. Clearly, as the model grows larger, one must deal with the problem of state explosion. That is another part of our ongoing research on multiresolution modeling and verification. The idea is for DEVS-Suite to choose the highest abstraction possible for verifying a certain property to reduce computation time.

\begin{table}
\centering
\caption{Sample runs of the circular buffer verification}
\label{table:circularBufferVerification}
    \begin{tabular}{| c | c | c | c | c |}
    \hline
 Buffer & Number of & State Space & Number of & Execution Time\\
 Size & Flit Types & Size & Cycles & (seconds)\\
   	\hline
   \hline
 8 & 25 & 3200 & 3328.6 & 4.55  \\[1mm]
 16 & 25 & 12600 & 13312.6 & 14.90  \\[1mm]
 32 & 100 & 201600 & 206848.6 & 273.78  \\[1mm]
	\hline
  \end{tabular}
\end{table}

We devised another example in \cite{gholami2017modeling} for minimal adaptive router. The model is specified in Constraint-DEVS and verified in DEVS-Suite. Also, it is worth noting that the state exploration algorithm and extend DEVS-Suite both support verifying coupled models as well. We have created and verified coupled models in this environment as well.

\section{Network-on-Chip Example Models}
\label{sec:nocExample}
In this section, we provide a model of Network-on-Chip (NoC) as a system that embodies more complex characteristics compared with the circular buffer. Here we intend to 1) detail the capabilities of the extended DEVS-Suite in simulating and model checking NoC models and 2) compare these capabilities with that of another popular modeling and verification environment (UPPAAL). While the circular buffer model (presented in this paper) provides a basic understanding of how Constraint-DEVS and DEVS-Suite operate, it is not intricate enough to demonstrate the capabilities of the DEVS-Suite framework.

Network-on-Chip \cite{hemani2000network} is a communication subsystem facilitating the interactions between intellectual properties on an SoC. There are various simulation tools suggested for NoC in various levels of abstraction such as BookSim \cite{jiang2013detailed}, GEM5 \cite{binkertgem5}, DARSIM \cite{lis2010darsim}, and Nostrum \cite{lu2005nnse}. Similarly, there has been research works on the verification of NoC, especially for properties such as deadlock, livelock, and packet delivery \cite{taktak2008tool,verbeek2010formal,salaun2007formal}. Performance-related properties are much more difficult to verify; this is due to the modeling constraints of these environments, the size of the state space, and limitations in property expressions.

As mentioned earlier, we compare DEVS-Suite with UPPAAL and not with any of the specialized NoC simulation or verification tools mentioned in the previous paragraph. There are two reasons for this: 1) none of the tools mentioned provide comprehensive modeling, validation, and verification tool for NoC and 2) none of the tools are generic: their support for validation or verification is dedicated to NoC. Both DEVS-Suite and UPPAAL, on the other hand, are generic tools supporting V\&V of any system as long as they are modeled with their supported modeling languages (namely DEVS and Timed Automata). Next, we first present the models of NoC in both of these environments and then compare them using a few representative experiments.

\subsection{NoC modeling: Timed Automata \& UPPAAL}
For the model of NoC in UPPAAL, we captured the creation of data in processing element (PE), packetization in network interface (NI), and routing in switch. We developed this model in Timed Automata (TA) and created several sample properties using temporal logic. The model is then implemented in UPPAAL (v. 4.0.14) and the properties are verified using the UPPAAL verifier.

\begin{figure}
    \centering
  	 \includegraphics[width=.8 \textwidth]{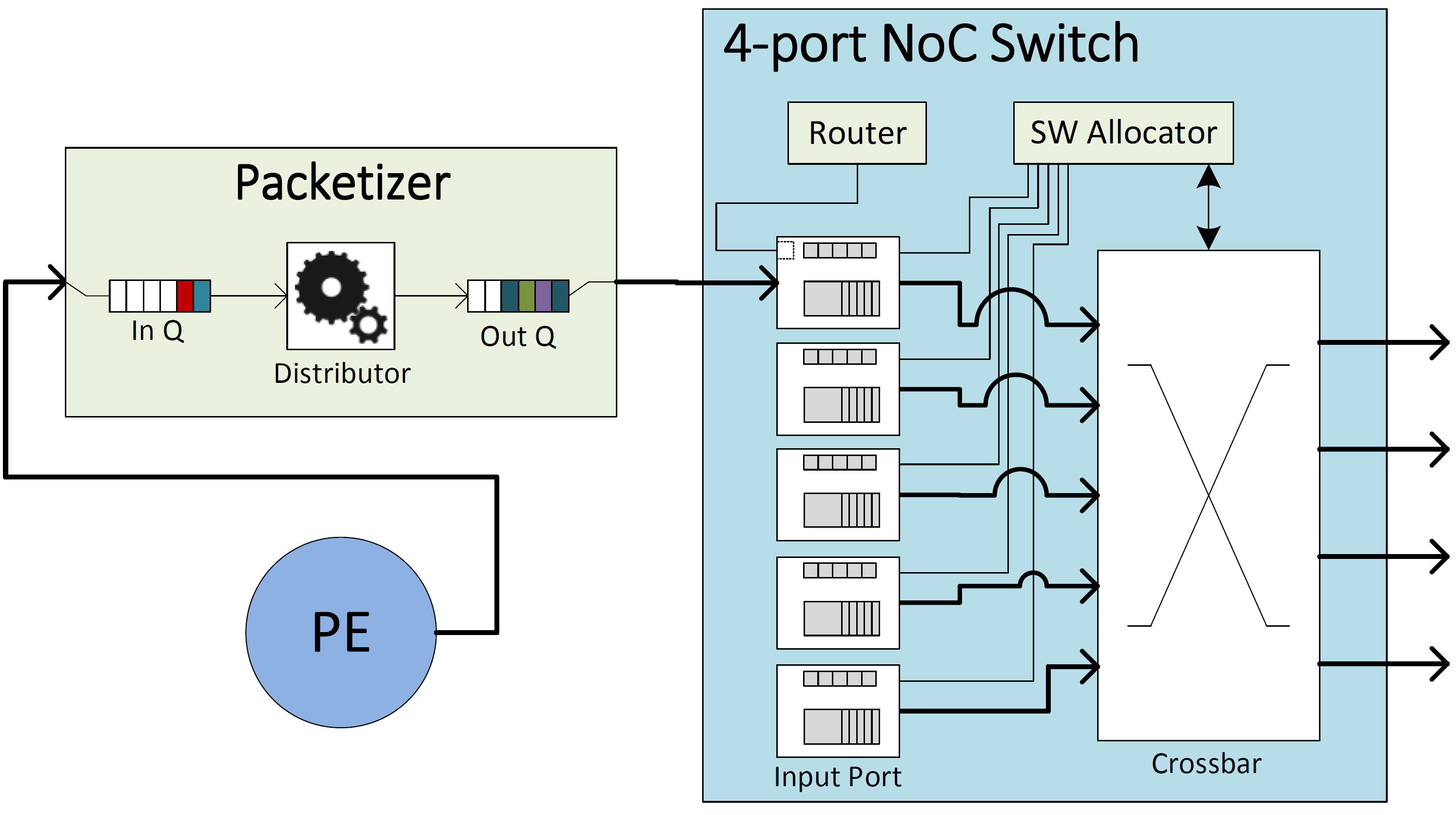}
    \caption{Detailed view of NoC as modeled in Timed Automata. Each component is a separate process.} \label{fig:uppaalModel}
\end{figure}

Data is first generated by the PE and sent to the Packetizer. The Packetizer stores the data and then converts it to packets in the order it receives them. Packets are then stored in a queue to be transmitted to the switch. Each packet, upon entering the switch from port $0$, goes through several steps to be transmitted out. First, it is stored in the input port along with other packets. Then, the router determines the suitable outgoing port for the packet based on the destination address stored in the header. Next, a path in the crossbar switch is allocated in order to transfer the packet from the input port to the output port. After that, the packet is transferred through the crossbar to the output port. Finally, the output port sends the packet out of the router on the link. The model of NoC developed is depicted in Figure \ref{fig:uppaalModel}.

Several simplifications are made for modeling NoC. First, the data which is transferred from PE to the switch is only a simple integer value. The value can be $0$, $1$, $2$, or $3$ corresponding to the output port that it must be sent to. Therefore, the Router component just reads the port it should forward the packet to. Also, the Distributor component in the Packetizer only forwards it to the output queue. However, both of these components are modeled and time is allocated to the tasks they run. Finally, other input ports in the switch are not connected to outside ports but generate random packets and allocate crossbar connections. This way contention is modeled and packets may be delayed. One important aspect of a switch that needs to be verified is the correctness of the switch allocation functionality.

\begin{figure}
    \centering
  	 \includegraphics[width=.9 \textwidth]{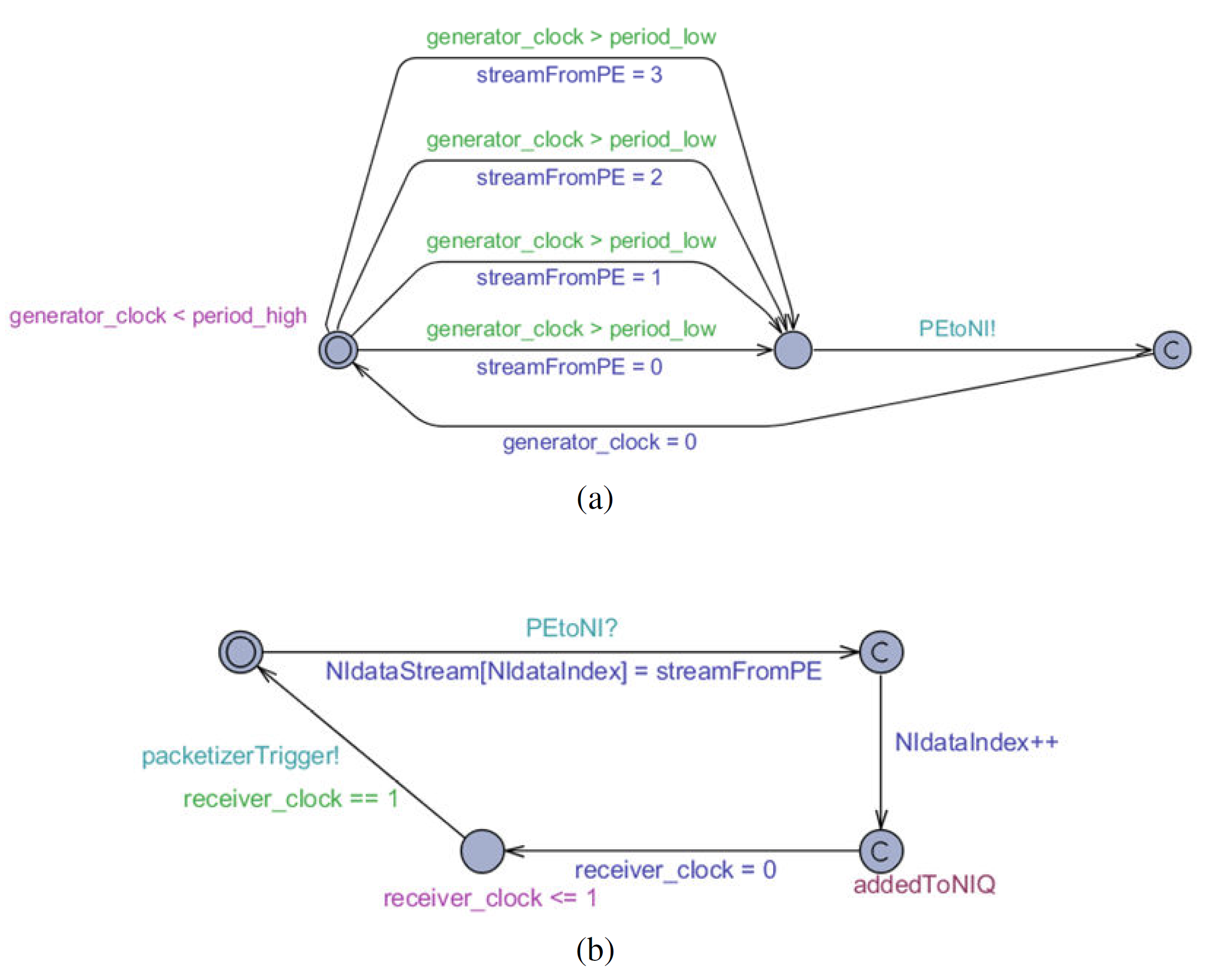}
    \caption{Two sample components of NoC and how they are connected via TA channels: (a) Task Generator (PE), (b) NI Receiver.} \label{fig:uppaalNoCTAModel2}
\end{figure}

Since these components are modeled as independent processes, a number of channels are introduced in Timed Automata for synchronization. Since information flow (i.e., there are no  messages to be transmitted) is not really possible in UPPAAL's Time Automata, global variables are defined and used to share data between two processes. Figure \ref{fig:uppaalNoCTAModel2} contains the TA model of two NoC components. The task generator (PE) and NI receiver. The task generator randomly creates new data streams to be sent to the network. A global variable (to communicate data) and a channel (for notification) are created to carry out the communication between these two components. The global variable, \emph{streamFromPE}, contains one of $0$, $1$, $2$, $3$ numbers. It is generated in the PE and later stored in a queue structure (\emph{NIdataStream}) in the NI receiver for packetization. The channel, named \emph{PEtoNI}, signals NI receiver that a new data stream is generated in PE and is stored in the global variable.

\begin{figure}
    \centering
  	 \includegraphics[width=.8 \textwidth]{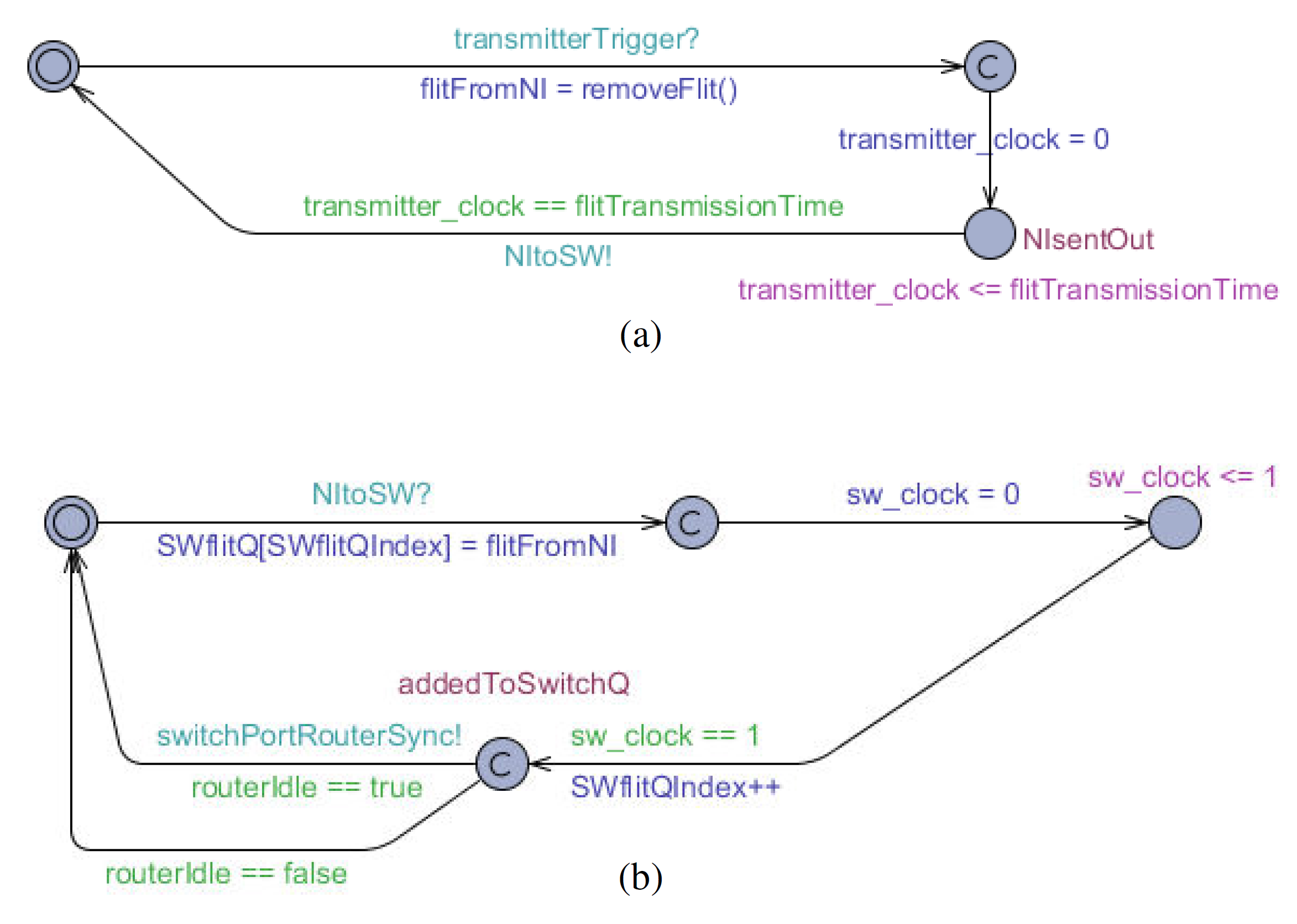}
    \caption{Two sample components of NoC and how they are connected via TA channels: (a) NI Transmitter, (b) Switch Input Port.} \label{fig:uppaalNoCTAModel}
\end{figure}

Another pair of components are illustrated in Figure \ref{fig:uppaalNoCTAModel}. In this figure, the two processes illustrated are the NI Transmitter and the Switch Input Port. Similar to the previous example, the communication between these processes is facilitated through a channel (\emph{NItoSW}) and a global variable (\emph{flitFromNI}). Other components of the system (as depicted in Figure \ref{fig:uppaalModel}) are modeled in UPPAAL and the communication between them is created using channels and global variables. 

Various properties can be defined for this NoC. We devised several desirable properties in three categories. For \textit{safety}, let us assume that a requirement of this NoC is not to drop packets. So, it would be a violation of safety if a full queue receives a new packet. For example, the index of the switch input queue must always be smaller than the length of the queue. This can be formulated using temporal logic in the following way: $\square(\mathit{SWFlitQIndex}<\mathit{switchQSize})$. As for \textit{liveness}, we formulate a property that ensures a packet entered a component eventually leaves it. For the network interface component, this property is written in temporal logic in the following way: $\square(\mathit{NI_Rec}.\mathit{addedToNIQ} \to \diamond \mathit{NI_Transmit}.\mathit{NIsentOut})$.

\subsection{NoC modeling: Constraint-DEVS \& DEVS-Suite}
\begin{figure}
    \centering
  	 \includegraphics[width=.85 \textwidth]{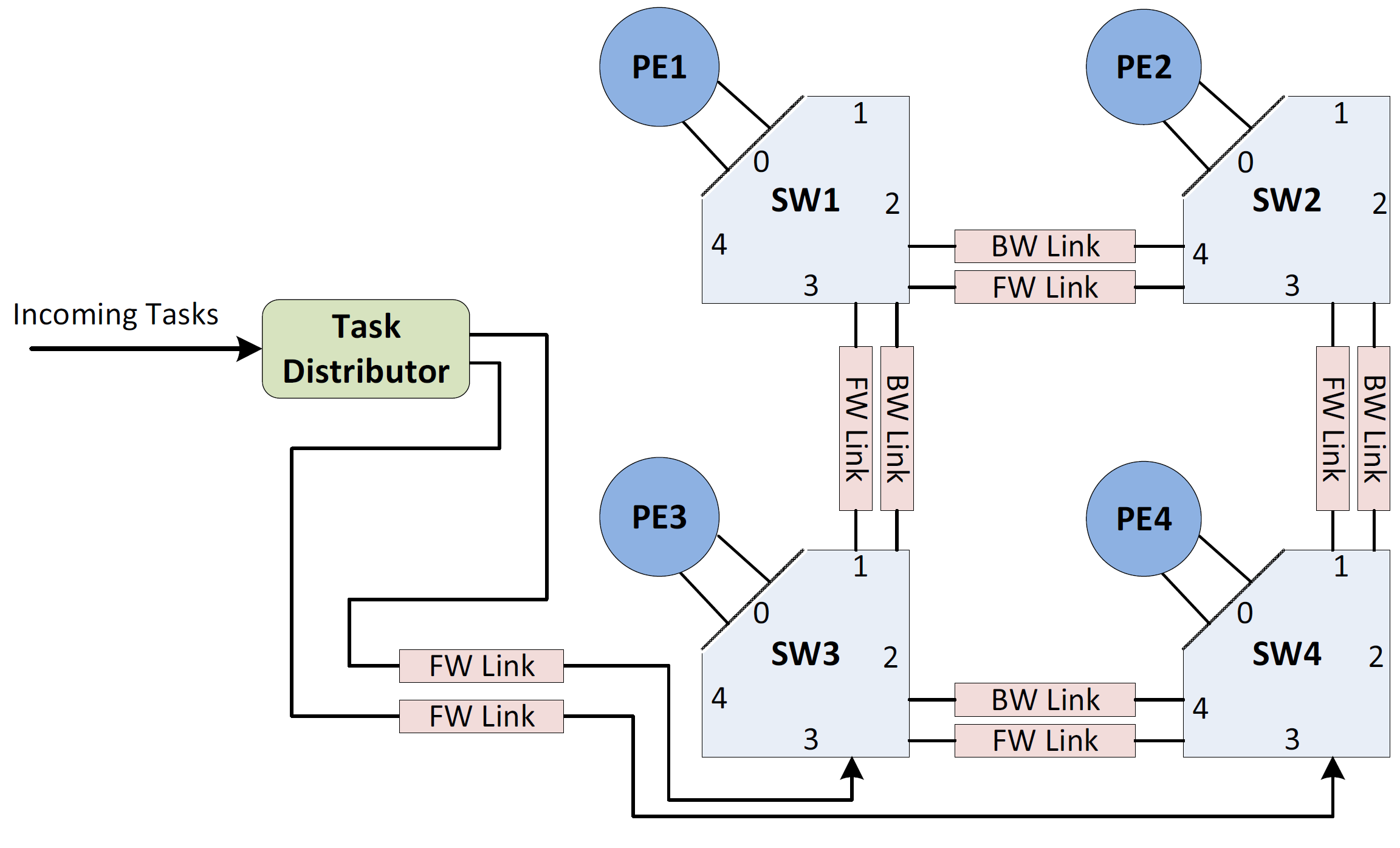}
    \caption{A 4-node model of NoC with the task distributor component.} \label{fig:nocWithTaskDistributorModel}
\end{figure}

We developed an NoC model using Constraint-DEVS and realized it in DEVS-Suite to showcase the additional capabilities provided to modelers in this environment. Contrary to the TA model, this NoC model has multiple nodes. One other difference here is that task generation is not random at the Processing Element (PE) level. Tasks come from outside and are distributed among PEs by a \emph{Task Distributor} component. As it is shown in the latter part of this section, a task distributor may have various policies in regard to task distribution. In this model of NoC, we connect the task distributor to the first row of switches (the southern edge). Figure \ref{fig:nocWithTaskDistributorModel} depicts a 4-node network with the task distributor.

Every task entering the network has a life cycle of 3; meaning it is processed by 3 different processing elements (one after another) before being considered complete. Every PE after receiving a task, processes it, increments its age, and then forwards it to the next PE using the NoC. PEs have different processing power while tasks have computation requirement. The processing time of a task in a PE is $t=\dfrac{\mathit{Computation \; Requirement}}{\mathit{Processing \; Power}}$.

\begin{figure}
    \centering
  	 \includegraphics[width=1 \textwidth]{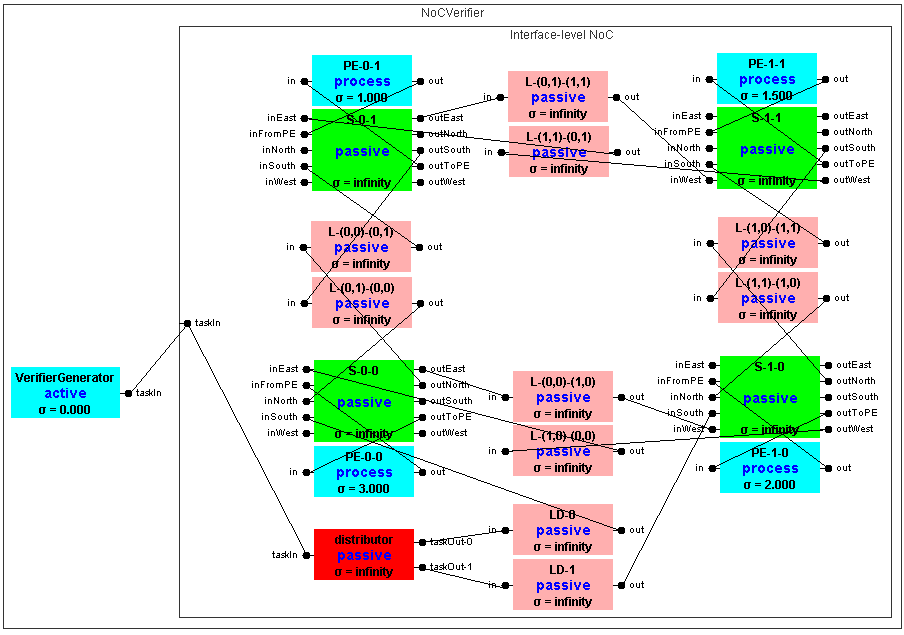}
    \caption{The verification environment for a 4-node model of NoC with major components: PE, Switch, Link, and Task Distributor in DEVS-Suite.} \label{fig:nocDEVSSuiteModel}
\end{figure}

The model of NoC developed in DEVS-Suite is suitable for both simulation and model checking. In the case of model checking, similar to the circular buffer, a \emph{Generator} is automatically generated to inject various combinations of inputs. A view of NoC with only 4 nodes (visualized in DEVS-Suite) is shown in Figure \ref{fig:nocDEVSSuiteModel}. We conduct experiments on larger networks for scalability evaluations (see Table~\ref{table:performanceTesting}). The verification engine explores the entire state space and constructs a reachability graph. The transducer is notified after each state change and it stores data within nodes of the reachability graph. The data stored by the transducer relevant to the properties it is verifying for that model. For the network created here, one can observe properties such as queuing times, component utilization, and performance. Upon having the entire reachability graph, the transducer can analyze the graph and verify various properties.

\subsection{Model Checking Experiments}
For both the DEVS-Suite and UPPAAL, there are some basic checks, including making sure no packets are dropped. We have developed two additional experiments for DEVS-Suite. One is intended to demonstrate DEVS-Suite's capability in verifying complex properties. For this purpose, we consider maximum and minimum PE utilization and the impact of task distribution on this. The second experiment is an observation on the performance of verification in DEVS-Suite relative to the size of state space.

\subsubsection{Experiment 1: PE Utilization and Task Distribution}
\label{sec:experiment1Description}
As explained earlier, upon receiving new tasks, the task distributor component decides which processing elements should process the task first. This decision can be guided by different policies. We developed two task distributors with different policies. One uses the round-robin approach and the other is adaptive. In the adaptive distribution policy, the distributor estimates the load on each processor by considering the history of distribution. Upon receiving a new task, it assigns it to the PE with the estimated lightest load.

We conduct verification of a 4-node NoC for each of these task distributions. At the end of the verification process, the transducer calculates the highest and lowest possible utilization for each PE by applying DFS (depth-first search) to the reachability graph. We evaluate the impact of this adaptive policy on utilization and whether it contributes to a more computationally balanced network.

For this experiment, we only change the task distributor component and the rest of the network remains unchanged. It is worth noting that the adaptive task distributor requires new \emph{explorable} state variables to hold its load estimates for PEs. We will analyze the impact of these new state variables on the size of the state space and the performance of the verification engine in Section \ref{sec:resultsAndAnalysis}. The execution platform for this experiment is a dual-core i5 3.5GHz machine running Java 7 and Windows 7.

\subsubsection{Experiment 2: Performance analysis and NoC Size}
For this experiment, we observe the performance of the DEVS-Suite verification engine for different network sizes. Here we are not verifying any particular property; however, we intend to explore the entire state space and show the practicality of applying the Constraint-DEVS model checking approach to a system such as NoC. We increase the size of the network and collect performance metrics from the verification engine.

The model of NoC, which is explored here, considers every state variable within the network as \emph{explorable} except for the processing queue of PE. We assume that any packet delivered to any PE will eventually be processed. Therefore, the phase, sigma, and buffer state variables for the switches, links, and PEs are all taken into account for model checking. As expected, the size of the state space rapidly grows and causes the familiar state explosion problem.

It should be noted that state exploration in DEVS-Suite is much more demanding as we are dealing with complex data types and, more generally, compound objects. Simpler modeling methods such as Timed Automata and Petri net may use primitive data types to represent the state of the system. However, in DEVS-Suite, in order to maintain the capability of simulation and experimentation, the models are more complex; thus, the exploration protocol, hashing method for visited and unvisited states, state modification, and model execution are all more demanding. This results in observing longer execution times for exploring the state spaces of models.

In this experiment, we will see the state explosion phenomenon for NoC models and present what DEVS-Suite and Constraint-DEVS can offer to make the verification of these complex models practical. Since this experiment is resource-intensive, we ran these experiments on a Windows 7 machine with quad-core Xeon 3.1GHz and 16GB of memory.

\subsection{Results \& Analysis}
\label{sec:resultsAndAnalysis}
For Experiment 1, the use of round-robin and adaptive distributors, as expected, show very different behaviors in the network. Table \ref{table:PEUtilizationTable} contains the maximum and minimum utilization of each processing element using the two distribution methods. These results are extracted from the reachability graph by applying DFS on every reachable path. As shown in the table, the adaptive approach results in a much more balanced load on the processing elements (looking at maximum or minimum columns for all processors).

As explained earlier (and shown in Figure \ref{fig:nocWithTaskDistributorModel}), the task distributor is only connected to the first row of switches (the southern edge). The round-robin approach not only takes into account the load on each processor, it also does not consider the time it takes for the task to be transported by the network to the PEs on upper rows. The adaptive distributor considers both of these and creates a more balanced distribution.

As described in Section \ref{sec:experiment1Description}, the adaptive distributor requires new \emph{explorable} state variables which increase the size of the state space. The last row in Table \ref{table:PEUtilizationTable} shows the duration of model checking of these 4-node networks with each of these distributors. The use of adaptive distribution increases the size of the state space by a factor of 15 which consequently increases the duration of the verification process as reported in Table \ref{table:PEUtilizationTable} (duration row).

\begin{table}
\centering
\caption{Maximum and minimum processor utilization using Round Robin (RR) and Adaptive task distributors with total execution time.}
\label{table:PEUtilizationTable}
\begin{tabular}{c|c|c||c|c|}
\cline{2-5}
& \multicolumn{2}{ c|| }{RR Distribution} & \multicolumn{2}{ c| }{Adaptive Distribution} \\ \cline{2-5}
& Min Util & Max Util & Min Util \; & Max Util \\ \cline{1-5}
\multicolumn{1}{ |c|| }{PE1} & 0.2 & 0.9 & 0.21 & 0.52 \\ \cline{1-5}
\multicolumn{1}{ |c|| }{PE2} & 0.1 & 0.46 & 0.09 & 0.56     \\ \cline{1-5}
\multicolumn{1}{ |c|| }{PE3} & 0.18 & 0.73 & 0.15 & 0.49      \\ \cline{1-5}
\multicolumn{1}{ |c|| }{PE4} & 0.15 & 0.49 & 0.15 & 0.54  \\ \cline{1-5}
\multicolumn{1}{ |c|| }{Overall} & 0.1 & 0.9 & 0.09 & 0.56  \\ \cline{1-5}
\multicolumn{1}{ |c|| }{Duration} & \multicolumn{2}{ c|| }{15.78s} & \multicolumn{2}{ c| }{245.20s} \\ \cline{1-5}
\end{tabular}
\end{table}

For Experiment 2, we increase the size of the network and explore the entire reachable state space. As expected, the size of the state space explodes as we add nodes to the network. The results are shown in Table \ref{table:performanceTesting}. It is clear that exploring the entire state space of the network as a whole can become computationally impractical. Of course, there is room for improving the performance of the engine by refactoring, creating lower-level code (such as in C), and using a faster platform. Still, one can easily see that scalability will always be a concern in exploring the state spaces of some systems.

The problem of state explosion is well-known and the results are not unexpected. So now the question is, what is the use of such a verification engine and is it practical to use this for a system such as NoC? This research shows that using this environment is beneficial and key for understanding, design, and developing large and complex systems. Below are four reasons why Constraint-DEVS and DEVS-Suite are useful for modeling complex systems:

\begin{description}
  \item[$\ast$ Partial Verification]: It may be unnecessary to explore the entire state space of a system such as NoC. We usually require model checking for small systems or portions of larger systems. For example, although model checking of networks larger than 16 nodes is impractical using basic computing platforms, model checking can be used to verify properties of individual components or a portion of the entire network.
  \item[$\ast$ Multiresolution Modeling]: Models created using Constraint-DEVS and DEVS-Suite can be created at various levels of abstraction. For the NoC example, we create models at different abstraction levels, each aiming at possessing certain details. Increasing the resolution (lower abstractions) results in models with larger state spaces. Some properties (such as routing-related properties, PE utilization, and latency) can be verified at higher abstraction levels without including all details. We have previously used MRM (Multiresolution Modeling) to create NoC models \cite{gholami2016multi}.
  \item[$\ast$ Validation \& Verification]: DEVS-Suite supports both validation (using simulation) and verification (using model checking) of Constraint-DEVS models. Models that lend themselves to model checking can be simulated without making any changes to them. This is beneficial for highly detailed models (high resolution) in which model checking a large portion is futile. Therefore, we can rely on a combination of small portion model checking and large scale simulation of high-resolution models.
  \item[$\ast$ Selective State Exploration]: State exploration is selective in DEVS-Suite as modelers can mark state variables as explorable/unexplorable based on the properties they intend to verify.
\end{description}

\begin{table}
\centering
\caption{Duration of model checking and size of the reachable state space for various network sizes.}
\label{table:performanceTesting}
    \begin{tabular}{| c | c | c |}
    \hline
 Network Size & Number of States & Duration\\
   	\hline
   \hline
 $2 \times 2$ & 18339 & 14.62s   \\[1mm]
 $3 \times 2$ & 108432 & 111.59s $\approx$ 2mins \\[1mm]
 $3 \times 3$ & 1748859 & 2029.11 $\approx$ 34mins  \\[1mm]
 $4 \times 3$ & 25255495 & 39211.6s $\approx$ 11hrs \\[1mm]
 $4 \times 4$ & -- & --  \\[1mm]
	\hline
  \end{tabular}
\end{table}

\subsubsection{Comparing UPPAAL and DEVS-Suite}
UPPAAL and DEVS-Suite offer certain distinct capabilities with respect to modeling, validation, and verification (see Table~\ref{table:comparisonTableForConstrainedDEVS}). We compare these using the following six criteria:

\begin{description}
  \item[$\ast$ Complex Data Transfer]: modeling time-sensitive information flow among modules is not supported in standard Time Automata \cite{lanotte2005timed}. In UPPAAL, complex data exchange is only possible through global variables and thus limited and not extendable. In the model presented above, global variables were used for communication and channels were utilized for synchronization. Complex data exchange is supported in DEVS and DEVS-Suite (in the form of Java Object).
  \item[$\ast$ Property Checking Capability]: UPPAAL, although being concise and efficient, is limited with respect to expressing QoS properties. An example is time-bound properties such as average queuing time for a sub-group of packets. This method of expressing properties is more efficient, but it is limited compared to DEVS-based property checking.
  \item[$\ast$ Structural Modeling]: UPPAAL does not intrinsically lend itself to modeling object-related, structural, or physical aspects of systems. It is intended for modeling functionality and logical data flow. Properties such as interfaces, input/output data types, and couplings with other components are missing from Timed Automata models. DEVS-Suite supports both structural and behavioral modeling.
  \item[$\ast$ Fine-grained Functional Modeling]: modeling of behavior in standard Timed Automata is fined-grained. Therefore, complex behavior (such as adaptive routing, flow control mechanisms, and arbitration) can be unnecessarily complicated. This shortcoming has been rectified in UPPAAL by providing Java programming language. Some aspects of the system which are too complex to be modeled using locations, edges, and signals can be implemented using Java programming language. The same thing applies to DEVS and DEVS-Suite; while DEVS formalism is limited in expressing fine-grain behavior, its extension Activity-based DEVS modeling \cite{alshareef2018activity} can be used. The realization of such fine-grain behavior specification is straightforward since the DEVS-Suite is implemented in the Java programming language.
  \item[$\ast$ Experimentation Capability for Simulator]: the simulator for UPPAAL is aimed at showing how the model operates through time. However, without proper monitoring and experimentation mechanisms, the simulation capability is difficult to use. The situation worsens if the model contains a number of modules operating in parallel. DEVS-Suite supports Parallel DEVS and simulation with built-in support for multi-mode superdense time trajectory tracking, animation, database repository, unit testing, and debugging.
  \item[$\ast$ Execution Performance]: UPPAAL provides fast simulation and model checking capability for TA models. The execution performance of DEVS-Suite simulation and model checking can rapidly degrade as the scale of the model grows. However, this is an important trade-off for the verification and validation of complex systems.
  \item[$\ast$ Selective State Exploration]: State variables in models created via Constraint-DEVS and DEVS-Suite can be selected instead of requiring developing different models having different state spaces. In any given model, each state variable can be considered for model checking (by marking it \emph{explorable}) or ignored by the verification engine (using \emph{unexplorable}). In contrast, every system modeled in UPPAAL has a fixed number of states for model checking unless a different model is created for it.
\end{description}

\section{Discussion}
\label{sec:discussion}
In this section, we briefly discuss the capabilities of the Constraint-DEVS approach to model checking and the extended DEVS-Suite framework. We compare the proposed work with other general formalisms/tools for modeling, validation, and verification (see Table \ref{table:comparisonTableForConstrainedDEVS}). Considering the focus on this research, we have not included other comparison criteria structural modeling (e.g., considering Petri net and Timed Automata), complex data transfer, or multi abstraction modeling. The Constraint-DEVS takes advantage of the DEVS system-theoretic and strong modularity and component-based modeling important for tackling scalability and complexity traits. This work is quite different from other DEVS-based approaches and tools (e.g., FD-DEVS and RTA-DEVS) because parallel DEVS modeling and simulation in DEVS-Suite is seamless and without having side effects on Constraint-DEVS and model checking.

\begin{table}
\centering
\caption{Comparing Constraint-DEVS with other well-known modeling methods.}
\label{table:comparisonTableForConstrainedDEVS}
    \begin{tabular}{| c | c | c | c | c | c | c |}
    \hline
 & \rotatebox[origin=c]{90}{\parbox[c]{2.5cm}{\centering Simulation}} & \rotatebox[origin=c]{90}{\parbox[c]{2.5cm}{\centering Verification}} & \rotatebox[origin=c]{90}{\parbox[c]{2.5cm}{\centering Formalism}}  & \rotatebox[origin=c]{90}{\parbox[c]{2.5cm}{\centering Non-determinism}} & \rotatebox[origin=c]{90}{\parbox[c]{2.5cm}{\centering Stochasticity}}  & \rotatebox[origin=c]{90}{\parbox[c]{2.5cm}{\centering Property Language}}\\
   	\hline
   \hline
 Timed Petri net & \checkmark & \checkmark & \checkmark & \checkmark & \checkmark & CTL  \\[1mm]
 Timed Automata & \checkmark & \checkmark & \checkmark & \checkmark & \checkmark & TCTL  \\[1mm]
 Constraint-DEVS & \checkmark & \checkmark & \checkmark & \checkmark & \checkmark & DEVS  \\[1mm]
 FD-DEVS & \checkmark & \checkmark & \checkmark & -- & -- & LTL  \\[1mm]
 PROMELA & -- & \checkmark & \checkmark & \checkmark & \checkmark & LTL  \\[1mm]
	\hline
  \end{tabular}
\end{table}

Table \ref{table:comparisonTableForDEVSSuite} compares extended DEVS-Suite with representative frameworks and tools supporting the modeling methods presented in Table 3. Some of these tools do not support all of the modeling, simulation, or verification. Therefore, they use other tools for missing capabilities such as model checking and designating states to be (un)explorable. For example, although MS4Me does not support model checking, the models created using it can be converted and used in SPIN.

\begin{table}
\centering
\caption{Comparing extended DEVS-Suite with a select number of verification common tools.}
\label{table:comparisonTableForDEVSSuite}
\begin{center}
    \begin{tabular}{| c | c | c | c | c | c | c |}
    \hline
 & \rotatebox[origin=c]{90}{\parbox[c]{2.5cm}{\centering Simulation}} & \rotatebox[origin=c]{90}{\parbox[c]{2.5cm}{\centering Verification}} & \rotatebox[origin=c]{90}{\parbox[c]{2.5cm}{\centering Modeling Environment}} & \rotatebox[origin=c]{90}{\parbox[c]{2.5cm}{\centering Simulation Environment}} & \rotatebox[origin=c]{90}{\parbox[c]{2.5cm}{\centering Verification Environment}} & \rotatebox[origin=c]{90}{\parbox[c]{2.5cm}{\centering Formalism}}\\
   	\hline
   \hline
 Extended DEVS-Suite & \checkmark & \checkmark & SELF & SELF & SELF & \checkmark  \\[1mm]
 UPPAAL & \checkmark & \checkmark & SELF & SELF & SELF & \checkmark  \\[1mm]
 Alpha/Sim & \checkmark & \checkmark & SELF & SELF & SELF & \checkmark  \\[1mm]
 SPIN & -- & \checkmark & -- & -- & SELF & --  \\[1mm]
 MS4ME & \checkmark & -- & SELF & SELF & -- & \checkmark  \\[1mm]
	\hline
  \end{tabular}
\end{center}
\end{table}

The advantages of extended DEVS-Suite is more evident when it is compared with domain specific tools for a specific system. We are currently using Constraint-DEVS and extended DEVS-Suite for NoC modeling, validation, and verification. Comparing this framework with existing NoC modeling tools produces more insight into the advantages and disadvantages of abstract modeling using Constraint-DEVS for simulation and model checking using DEVS-Suite.

\section{Conclusion}
\label{sec:conclusion}
In this paper, we presented the Constraint-DEVS as a DEVS-based modeling checking method. The extended DEVS-Suite framework supports the implementations of both DEVS and Constraint-DEVS specifications with simulation and model checking, respectively. Also, models of Network-on-Chip were created using Constraint-DEVS and verified using the extended DEVS-Suite framework. The models were used to show the significance of defining property expressions, property checking, data collection, and experimentation. The DEVS-Suite framework is capable of verifying path-based properties (properties relating to the state of the system over a prolonged period of time) as well as state-based properties.

Extending Parallel DEVS with Constraint-DEVS eliminates the need to manually convert models that are developed for simulation to other (usually less descriptive) models that support model checking. The process of converting current Parallel DEVS models to Constraint-DEVS models entails placing bounds on the values state variables using prescribed template atomic and coupled models. Switching between simulation and model checking is made possible by toggling a variable through simple model configuration. As mentioned earlier, the state exploration protocol in DEVS-Suite incorporates existing simulation protocol. It should be noted that the core architectural design and modularity of the DEVS-Suite played a major role in extending it to support model checking without disrupting any of simulation modeling and visualization capabilities.

Our ongoing research and goal include developing models of NoC at flit-level and realize them in DEVS-Suite. These models may be validated and verified for their desired behavioral properties in DEVS-Suite. Flit-level models expand the use of the verification engine to verify properties of NoC. Creating NoC models at flit level with virtual channels, flits, packetization/depacketization, and flow control mechanism unfold new properties to verify such as fairness, deadlock avoidance, minimum queue utilization, and load-balanced routing.

\bibliographystyle{ACM-Reference-Format}
\bibliography{references}

\appendix

\section{Verification in DEVS-Suite Framework}
\label{sec:verificationInDEVSSuite}
DEVS-Suite is extended to accommodate model checking by 1) developing Constraint-DEVS models and 2) their executions using the state exploration protocol provided in Listing \ref{state-exploration-protocol}. These two extensions are discussed in Sections \ref{sec:constrainedDEVSinDEVSSuite} and \ref{sec:explorationProtocol}.

\begin{figure}[h]
    \centering
  	 \includegraphics[width=.8 \textwidth]{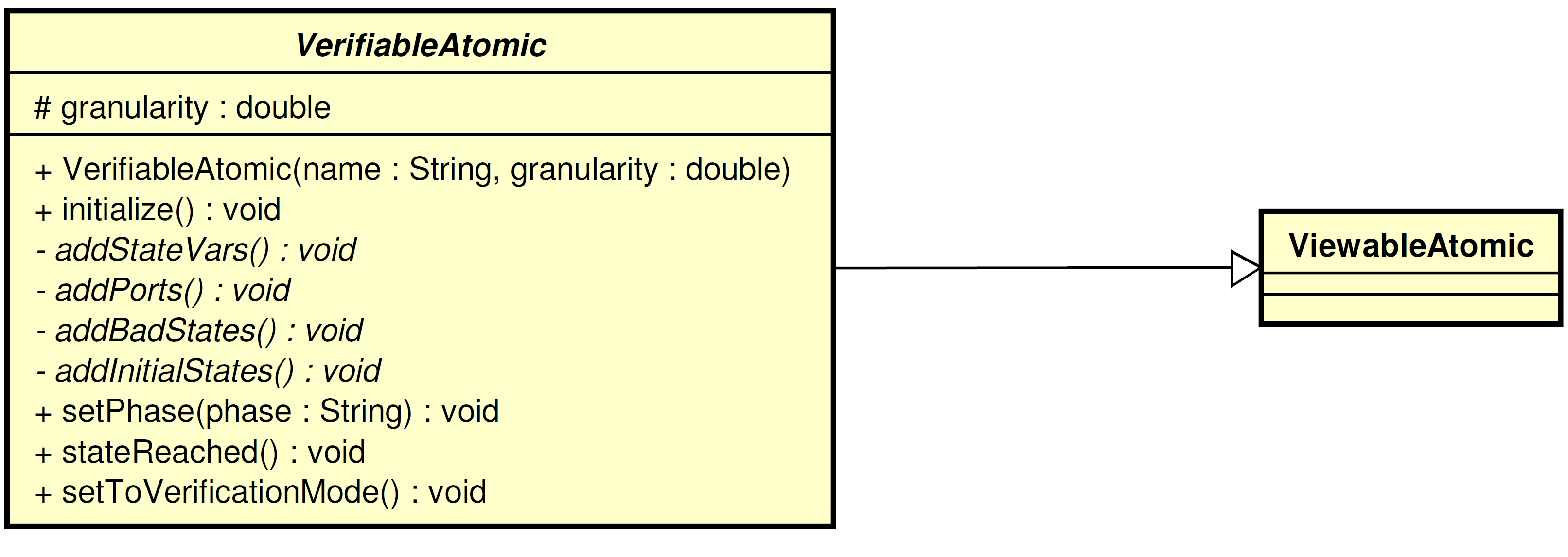}
    \caption{\textit{VerifiableAtomic} class for verification and validation.} \label{fig:verifiableAtomic}
\end{figure}

\subsection{Constraint-DEVS}
\label{sec:constrainedDEVSinDEVSSuite}
At the modeling level, base classes are added to support Constraint-DEVS. At the atomic level, DEVS atomic models are represented in DEVS-Suite with a Java class called \texttt{ViewableAtomic}. It has predefined methods for external, internal, and confluent transition functions and time advance. A new base class called \texttt{VerifiableAtomic} is implemented which extends \texttt{ViewableAtomic} with some required verification means such as adding initial states, invalid states, and input/output ports. The class is shown in Figure \ref{fig:verifiableAtomic}. This summarized view of \texttt{VerifiableAtomic} also shows additional functionality for state transition checking (\texttt{stateReached}), initialization of state variables (\texttt{initialize}), and switching between simulation and model checking modes (\texttt{setToVerificationMode}).

\begin{figure}
    \centering
  	 \includegraphics[width=1 \textwidth]{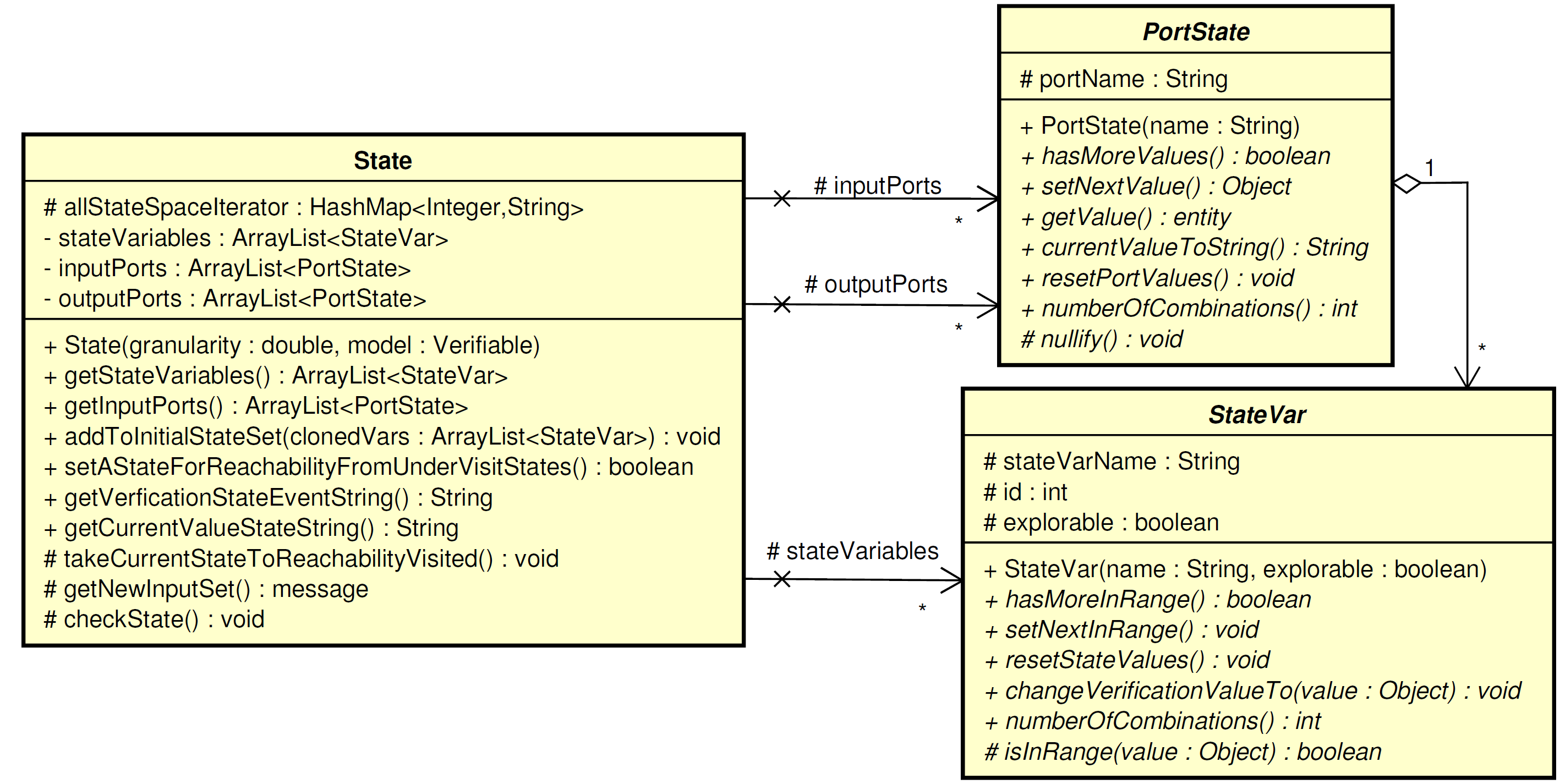}
    \caption{\emph{State}, \emph{StateVar}, and \emph{PortState} classes for state space exploration.} \label{fig:state-stateVar-uml}
\end{figure}

This base class is capable of holding bounded state variables (defined by regular expressions) and using them in model execution. However, DEVS-Suite does not have support for bounded state variables. Therefore, some extensions are made. Figure \ref{fig:state-stateVar-uml} contains 3 classes necessary for state exploration. The \texttt{StateVar} abstract class is the basis for any state variable; in other words, any state variable that is required to be explored during model checking has to extend this class. Such bounded state variables are necessary for state exploration. Similarly, external ports should extend the \texttt{PortState} class. Similar to the \texttt{StateVar}, \texttt{PortState} implements bounded incoming ports. Instances of \texttt{message} class are communicated between models using ports. Each instance of message may contain a number of \texttt{content} objects which are the actual data communicated between two models. \texttt{PortState} ensures values taken by the \texttt{content} object are bounded. It is used by the verification engine to create all possible external events and injecting them into the model. \texttt{PortState} may contain one or more of \texttt{StateVar} instances for modeling bounded port values.

Another base class in relation to state variables is the \emph{State} class. While a model may have a number of \texttt{StateVar} and \texttt{PortState} instances (for multiple state variables and input/output ports respectively), it can only have one one instance of \emph{State}. The \texttt{State} class holds all state variables (instances of \texttt{StateVar} class), all ports and their possible values (instances of \texttt{PortState}), and data structures for visited and unvisited states (during state exploration). Figure \ref{fig:state-stateVar-uml} depicts the relationship among \texttt{State}, \texttt{StateVar}, and \texttt{PortState} classes.

The classes in Figures \ref{fig:verifiableAtomic} and \ref{fig:state-stateVar-uml} show only some of their attributes and operations; including all their attributes, methods, and initializations is impractical. The functionalities left in the UML class diagram are related to the exploration functionality. More complete UML class diagrams, but without details of the classifiers, are provided in Section~\ref{sec:explorationProtocol}.

\subsection{State Space Exploration Protocol}
\label{sec:explorationProtocol}
With the Constraint-DEVS modeling added to DEVS-Suite, next the model checking algorithm presented in Listing \ref{state-exploration-protocol} must be added. The model checking algorithm uses the additional information provided in the models to conduct reachability analysis on the state space. The algorithm, with its main design concept and implementation, is highlighted here.

\begin{figure}
    \centering
  	 \includegraphics[width=1 \textwidth]{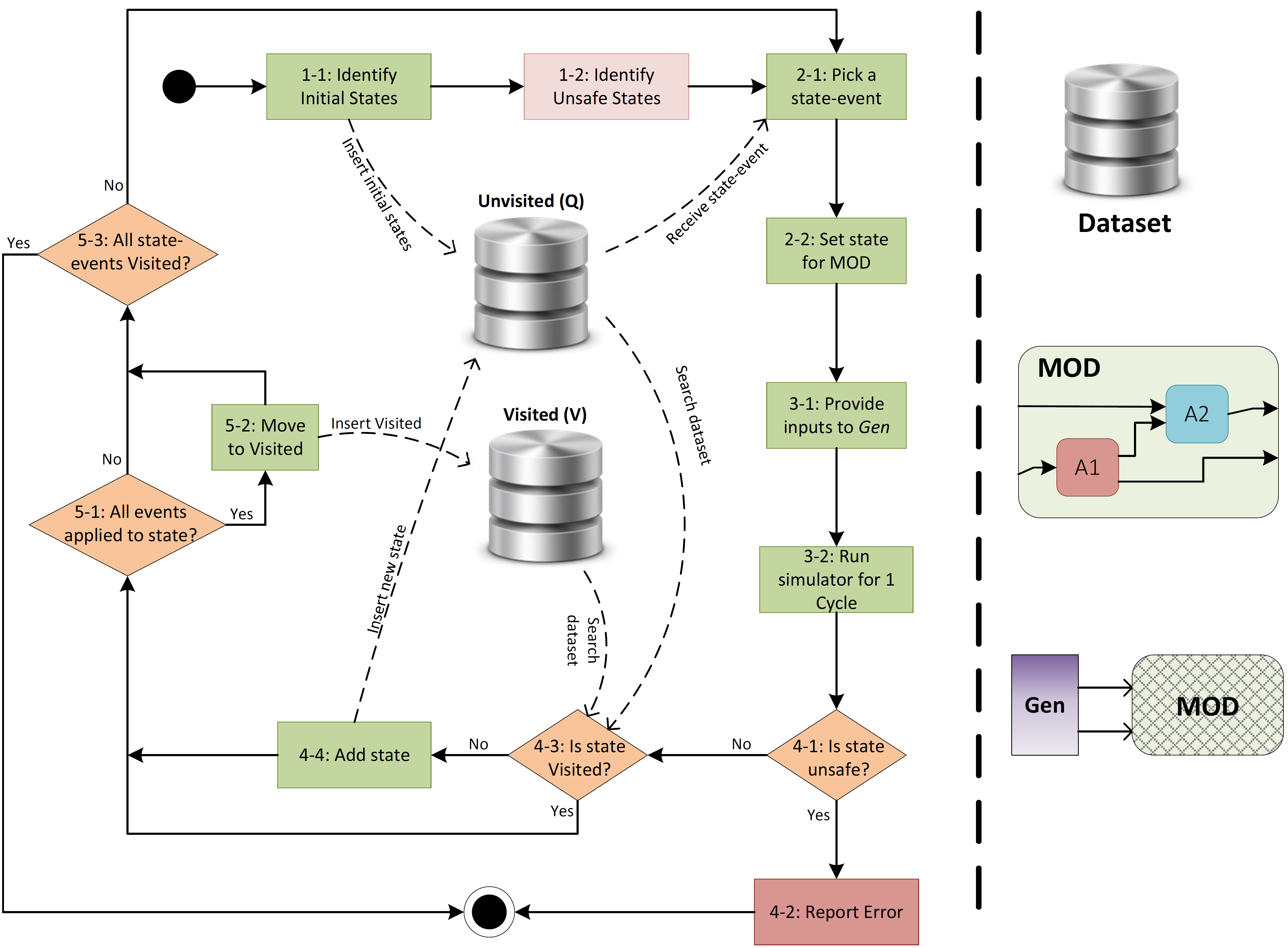}
    \caption{Flowchart for the Constraint-DEVS model checking protocol in DEVS-Suite.} \label{fig:protocolVisual}
\end{figure}

Before delving into extending the DEVS-Suite with the execution of Constraint-DEVS models, major parts in the state exploration task are described. There following are needed for realizing the model checking protocol illustrated in Figure \ref{fig:protocolVisual}:

\begin{description}
  \item[$\ast$ Verification Engine]: this is the module in charge of state exploration. It runs the protocol presented in Algorithm 1. Starting/ending the process is also the responsibility of this module. As explained earlier, the \emph{Verification Engine} uses the simulation protocol to carry out the state space exploration.
  \item[$\ast$ Target Model]: MOD is the target model to which the reachability analysis is applied to.
  \item[$\ast$ Experimental Frame]: the \emph{Generator} and \emph{Transducer} pair which are in charge of generating possible combinations of input and collecting the state transition trace, respectively.
  \item[$\ast$ State Sets]: \emph{Q} and \emph{V} are required for keeping track of visited and unvisited states.
\end{description}

The steps with decision nodes shown in Figure \ref{fig:protocolVisual} emphasizes state exploration. The flowchart at the left-hand side of this figure visualizes the model checking protocol provided earlier in Listing \ref{state-exploration-protocol}. The right-hand side provides the reader with a guide on the elements used in the flowchart. There are two datasets (Unvisited and Visited), MOD (the model under verification which could be atomic or coupled) and Gen (a generator that is automatically instantiated to inject inputs to MOD's input ports). There are 5 phases for the model checking protocol:

\begin{enumerate}
\item \textbf{Initialization:} during steps 1-1 and 1-2, initial states and unsafe states are identified and the initial ones are added to \emph{Q}. Other steps during initialization are initializations of the EF (\emph{Generator} and \emph{Transducer}) and the \emph{Verification Engine}.
\item \textbf{Main Loop (new state):} during steps 2-1 and 2-2, a state ($S_\mathit{current}$) is removed from \emph{Q} and is set by the \emph{Verification Engine} as the current state of MOD.
\item \textbf{Inner Loop (input injection):} during steps 3-1 and 3-2, a combination of possible input set is applied to MOD and the simulation engine is called to execute the model for a single cycle.
\item \textbf{Housekeeping:} at steps 4-1 up to 4-4, the resulting state is stored in the \emph{Transducer}, the unvisited ones in \emph{Q}, and if all input injections are applied to state $S_\mathit{current}$, that state is put into set \emph{V}.
\item \textbf{Termination:} at steps 5-1 up to 5-3, the process examines termination conditions. The process continues at the \emph{Main Loop} phase if \emph{Q} is not empty. Otherwise, it proceeds to the final state and the Transducer provides the user with trace and property checking results.
\end{enumerate}

\begin{figure}
    \centering
  	 \includegraphics[width=1 \textwidth]{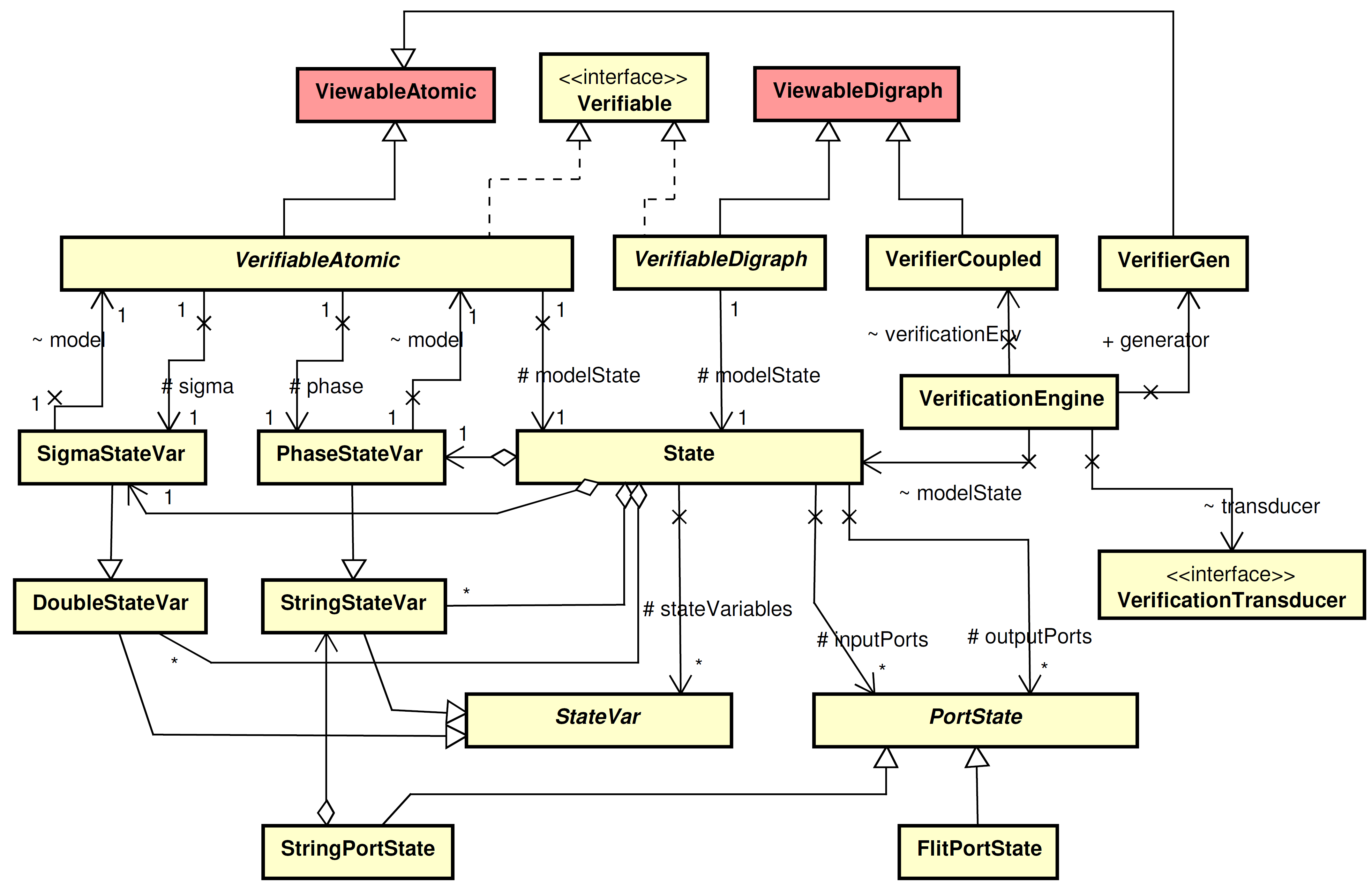}
    \caption{Partial UML diagram of DEVS-Suite modeling class structure (new classes in yellow.)} \label{fig:modelingPackageUML}
\end{figure}

This protocol required adding some extensions to DEVS-Suite. These along with their relationship with classes (e.g., \texttt{StateVar} and \texttt{State}) for Constraint-DEVS are shown as a UML class diagram (see Figure \ref{fig:modelingPackageUML}). This partial class diagram highlights some of the classifiers and their relationships to support both simulation and model checking.

As explained earlier, the state variables should all be of type \texttt{StateVar}. This class mandates implementing bounds on the state variable along with an iteration capability (so that all values can be iterated). Two examples of state variables are provided in Figure \ref{fig:modelingPackageUML}: \texttt{DoubleStateVar} for double state variables and \texttt{StringStateVar} for string state variables. Other state variables can be added as long as they are inherited from the \texttt{StateVar}. Phase and Sigma (required for model checking and simulation) extend \texttt{StringStateVar} and \texttt{DoubleStateVar}, respectively.

The \texttt{State} class holds all state variables and belongs to all instances of \texttt{VerifiableAtomic} and \texttt{VerifiableDigraph}. The model under test can be an atomic or a coupled model which both implement the \texttt{Verifiable} interface.

Finally, the \texttt{VerificationEngine} class is instantiated for model checking scenarios and manages the entire process of verification. It instantiates the \texttt{VerifierGen} class which produces all possible combinations of input values. In addition, it creates one or more instances of the Transducer class which are in charge of data collection, state-based analysis, and output validation. Finally, the responsibility of keeping track of visited and unvisited states and setting them (manually to the State class) for the model under test lies with \texttt{VerificationEngine}.

\begin{figure}
    \centering
  	 \includegraphics[width=.85 \textwidth]{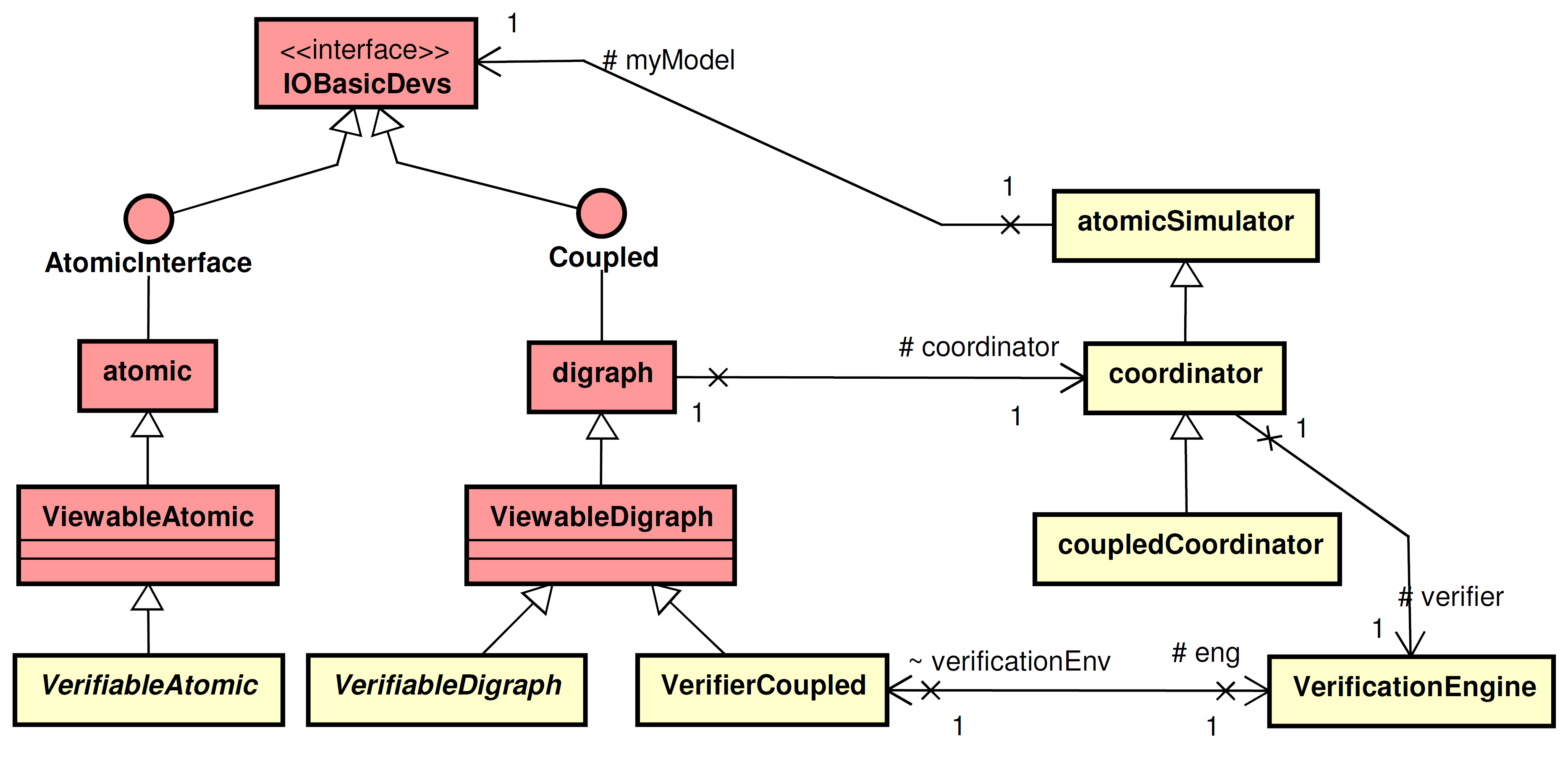}
    \caption{Partial UML diagram of DEVS-Suite model, simulation, and verifier packages (new and modified classes are in in color yellow.)} \label{fig:executionPackageUML}
\end{figure}

To keep the simulation functionality intact, we reuse DEVS-Suite atomic and coupled simulation capability. The \texttt{VerificationEngine} operates on top of these classes to facilitate model checking. Figure \ref{fig:executionPackageUML} provides an overview of the simulation package and how it relates to the modeling package. The bottom row of classes in the UML class diagram (plus minor changes without side effect in the \texttt{Coordinator} and \texttt{atomicSimulator} classes) paved the way for the added model checking functionality.

In order to reuse DEVS-Suite's animation and tracking capability, as illustrated in Figure \ref{fig:executionPackageUML}, the \texttt{VerifiableAtomic} and \texttt{VerifiableDigraph} classes extend \texttt{ViewableAtomic} and \texttt{ViewableDigraph} classes, respectively that support animation as well as linear and superdense time trajectory generation and viewing at run-time. Also, this relationship facilitates reusing the simulation capability without requiring any changes to the Constraint-DEVS models.

The operation of the state exploration protocol as presented in Listing \ref{state-exploration-protocol}, is developed in DEVS-Suite. The execution of the verification protocol contains hundreds of calls and interactions. Here we present the reader with a statechart and a partial sequence diagram.

\begin{figure}
    \centering
  	 \includegraphics[width=1 \textwidth]{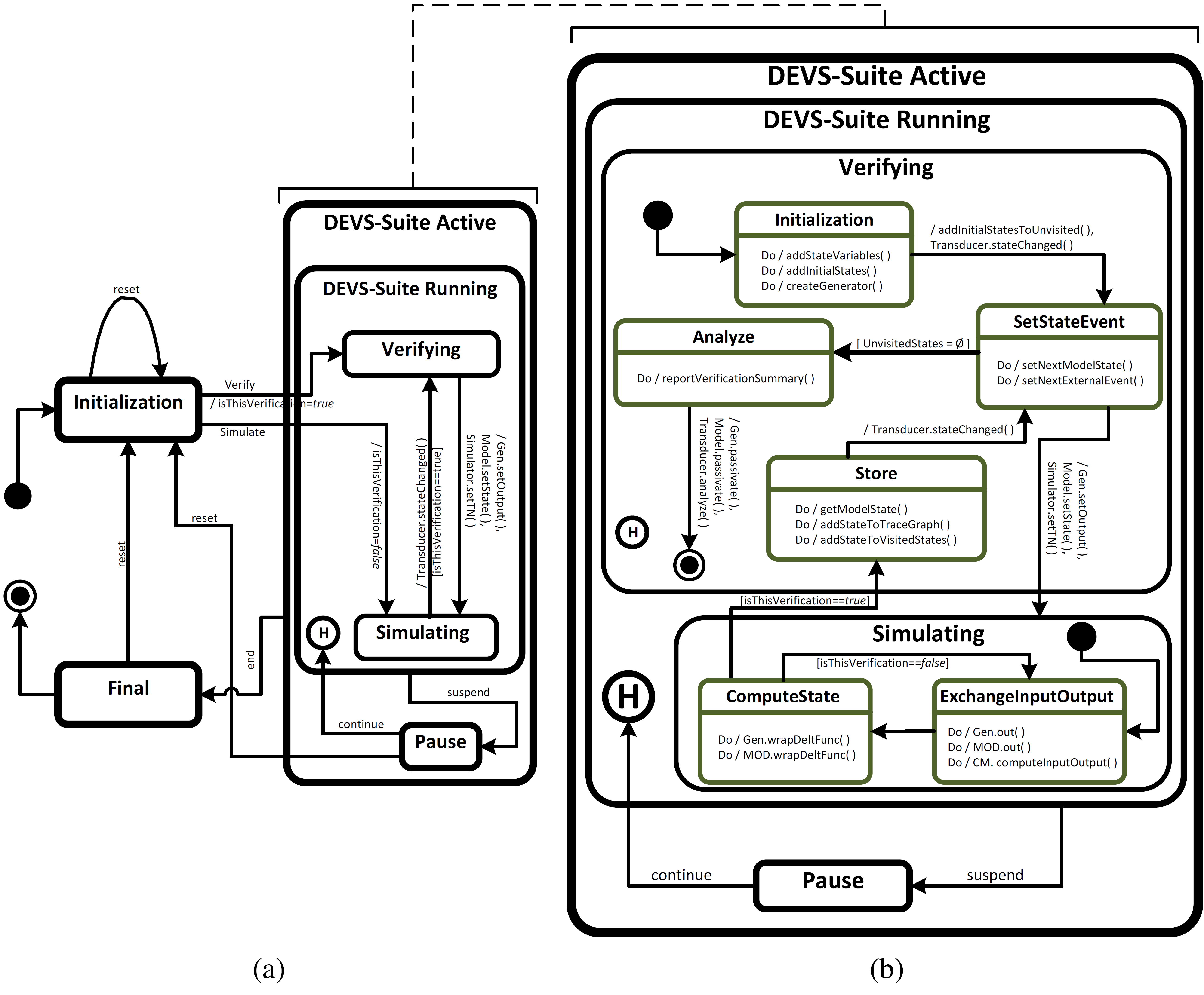}
    \caption{DEVS-Suite V\&V statechart: (a) Partial statechart diagram for DEVS-Suite from initialization to termination, (b) detailed view of the ``DEVS-Suite Active'' state (containing verification and simulation sub-states.)} \label{fig:protocolStatechartDiagram}
\end{figure}

In the Statecharts diagram presented in Figure \ref{fig:protocolStatechartDiagram}-a we show three major states in which DEVS-Suite may be in: \emph{Initialization}, \emph{Active}, and \emph{Final}. To break it down further, an active DEVS-Suite may be in \emph{Running} or in \emph{Pause} mode and while in Running mode, DEVS-Suite is either in \emph{Verifying} or \emph{Simulating} mode. Transitions between these modes (in Figure \ref{fig:protocolStatechartDiagram}-a) are all labeled with the signals that cause them and the actions that are executed due to them. Whether a model is being verified or simulated in the \emph{Running} mode is decided at \emph{Initialization}. The \emph{simulate} or \emph{verify} signals decide in which execution mode the model can be in. Other examples of signals are the \emph{reset} and \emph{suspend}. The \emph{reset} signal (caused by a user action) always takes the model back to the Initialization state. Similarly, while in \emph{Running} state (verifying or simulating a model), a \emph{suspend} signal takes DEVS-Suite to \emph{Pause} state. The \emph{continue} signal brings the model back to the \emph{Running} mode.

in Figure \ref{fig:protocolStatechartDiagram}-b, we provide a detailed view of the \emph{Active} state in which \emph{Verifying} and \emph{Simulating} compound states are shown in detail. The \emph{Verifying} compound state shows the phases that the verification engine goes through and the calls it makes to other processes. The process shown here is consistent with the exploration protocol in Listing \ref{state-exploration-protocol} and its visualization in Figure \ref{fig:protocolVisual}. As explained earlier, the verification engine uses the simulation engine to run the model. This is evident here with the transition between the \emph{SetStateEvent} (in \emph{Verifying}) to the \emph{Simulating} state and back. The verification engine picks a state event in \emph{SetStateEvent}, the simulation engine runs the model for one cycle (in \emph{Simulating}), and then the state is store in the suitable dataset (Visited or Unvisited) in the \emph{Store} state.

The \emph{suspend} signal brings DEVS-Suite to the \emph{Pause} state from \emph{Running}. In order to go back to the right state (depending on whether we were verifying or simulating the model) when the \emph{continue} signal is received, a history state is needed within the \emph{Running} compound mode.

\begin{landscape}
\begin{figure}
    \includegraphics[width=1\linewidth]{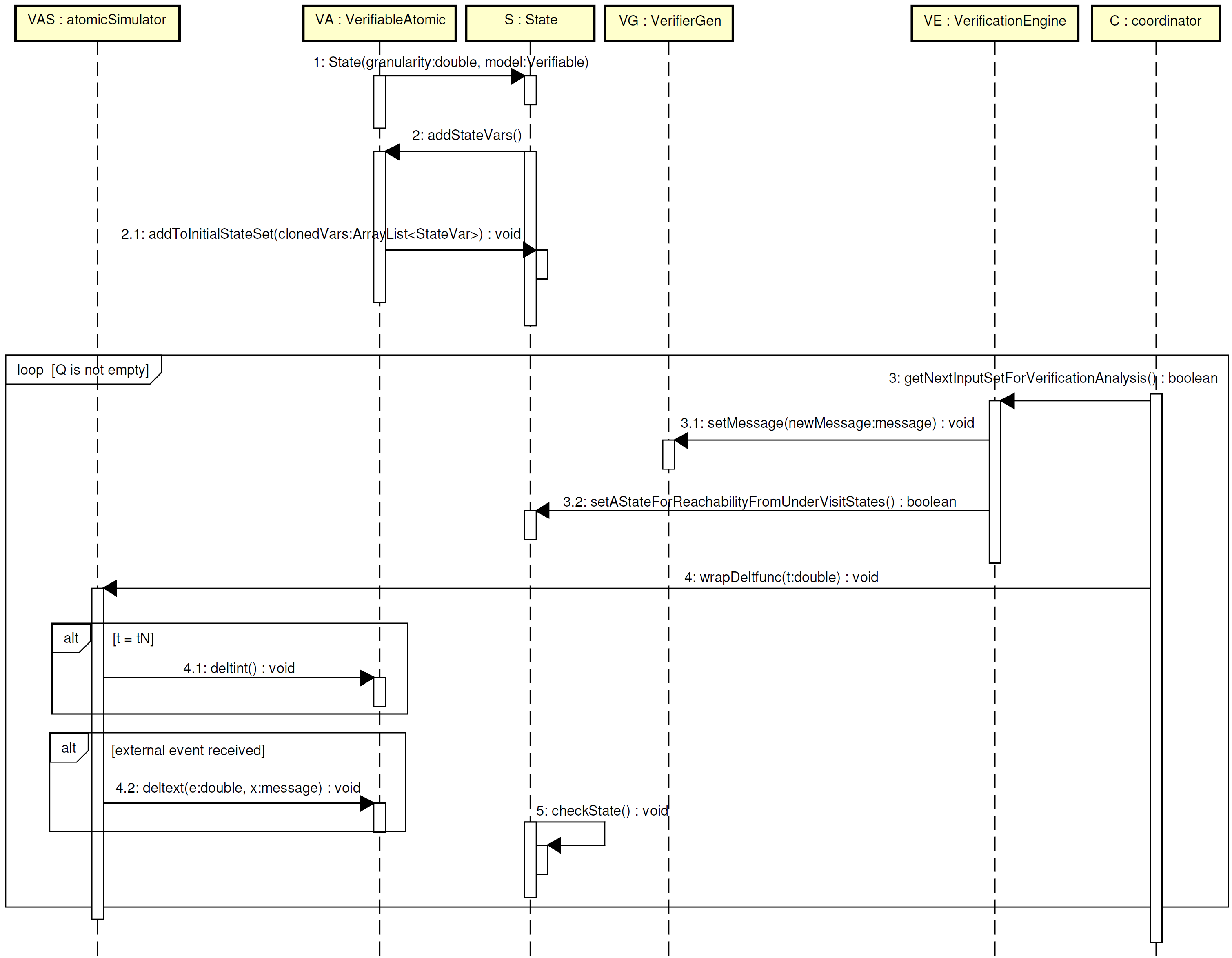}
    \caption{Partial UML sequence diagram exemplifying the execution of DEVS-Suite state space exploration protocol.} \label{fig:sequenceDiagram}
\end{figure}
\end{landscape}

The sequence diagram in Figure \ref{fig:sequenceDiagram} focuses on six major classes of the exploration protocol: \texttt{VerifiableAtomic} (MOD), \texttt{State} (containing the state variables and ports of MOD), \texttt{VerifierGen} (the generator of EF), \texttt{VerificationEngine} (orchestrator of the exploration process), and the two simulation classes uses by the verification engine: \texttt{atomicSimulator} and \texttt{coordinator}. Many objects and transitions within and among them that participate in a full cycle of state exploration must obviously excluded. The sequence diagram presented in Figure \ref{fig:sequenceDiagram}, contains the initialization and the exploration phases.

After the initialization phase is completed (state variables are identified, initialized, and put into the right datasets), the \texttt{coordinator} invokes the \texttt{VerificationEngine} at every cycle to perform state exploration. The \texttt{VerificationEngine} sets a new state for MOD and selects a set of external events to be injected by the \texttt{VerifierGen}. The \texttt{coordinator} then kick starts a single-cycle simulation by calling on the \texttt{atomicSimualtor}. The \texttt{atomicSimulator} by itself calls the MOD to react to the external events and passage of time. Finally, the new state of the MOD is analyzed and added to the datasets for further analysis. This process is repeated until there are more states to visit.

\end{document}